%% file: main.tex
\pdfoutput=1
\documentclass[12pt]{article}

\newif\ifsubmission
\submissionfalse   
\newif\ifpreview
\previewfalse   

\input{preamble}

\usepackage{natbib}
\usepackage[affil-it]{authblk}
\usepackage{amsthm}
\usepackage{float}
\usepackage{enumitem}
\newtheorem{prop}{Proposition}

\newtcolorbox{mybox}{
  colback=gray!5,
  colframe=black,
  boxrule=0.5pt,
  arc=2pt
}
\newtcolorbox{titlebox}[1][]{
  colback=gray!5,
  colframe=gray!150,
  fonttitle=\bfseries,
  title=#1,
  boxrule=0.8pt,
  arc=3pt,
}
\newcommand{\neweqbox}[2]{%
  \newtcolorbox{#1}{%
    enhanced,
    colback=#2!8, colframe=#2!45,
    boxrule=0.4pt, arc=2pt,
    left=4pt, right=4pt, top=2pt, bottom=2pt,
    before skip=2mm, after skip=2mm,
    before upper={\setlength{\abovedisplayskip}{0pt}\setlength{\abovedisplayshortskip}{0pt}%
                  \setlength{\belowdisplayskip}{0pt}\setlength{\belowdisplayshortskip}{0pt}}}}
\neweqbox{eqshadeNBD}{BurntOrange}
\neweqbox{eqshadeGGG}{BrickRed}

\newtcolorbox{modelbox}[3][gray]{
  enhanced,
  colback=#1!5,
  colframe=#1!60!black,
  boxrule=0.8pt,
  arc=3pt,
  left=8pt, right=8pt, top=4pt, bottom=4pt,
  before upper={\linespread{1}\selectfont%
                \setlength{\abovedisplayskip}{0pt}\setlength{\abovedisplayshortskip}{0pt}%
                \setlength{\belowdisplayskip}{0pt}\setlength{\belowdisplayshortskip}{0pt}},
  overlay unbroken and first={
    \node[anchor=west, font=\scriptsize\sffamily\bfseries,
          text=#1!60!black, fill=#1!5, inner xsep=5pt, inner ysep=1.5pt]
      at ([xshift=10pt, yshift=-2.5pt]frame.north west) {\MakeUppercase{#2}};
    \node[anchor=east, font=\scriptsize\sffamily\bfseries,
          text=#1!60!black, fill=#1!5, draw=#1!60!black, line width=0.6pt,
          rounded corners=4pt, inner xsep=6pt, inner ysep=2.5pt]
      at ([xshift=-10pt, yshift=-2.5pt]frame.north east) {#3};
  },
}

\input{macros.tex}

\graphicspath{{./}{figures/}}

\begin{document}

\ifsubmission
\begin{center}
  {\LARGE\bfseries Scalable Amortized Variational Inference\\[0.3em]
   for Non-Poisson Buy-`Til-You-Die Models\par}
\end{center}
\vspace{1.5em}
\else
\title{Scalable Amortized Variational Inference\\for Non-Poisson Buy-`Til-You-Die Models}
\author[1]{Sulagna Ghosh}
\author[2]{Aaron Schein}
\affil[1]{Department of Statistics, The University of Chicago}
\affil[2]{Data Science Institute, The University of Chicago}  
\maketitle
\fi

\input{sections/0_abstract}
\input{sections/1_intro}

\input{sections/2_data}

\input{sections/3_background}

\input{sections/4_model}

\input{sections/5_inference}
\input{sections/6_simulations}

\input{sections/7_case_study}

\input{sections/8_conclusion}

\ifsubmission\else
\section{Acknowledgements}
 AS helped develop a precursor to this work called ``Py(Torch) `Til You Die'' while he was a research scientist 2020--2022 at Ocurate (since acquired by Fenix Commerce), during which time he benefitted greatly from close collaboration and support from Gio Fernandes, Harlan Holt, Vivian Ellis, Sabrina Margetic, David Rothschild, and Tobias Konitzer, among others at Ocurate, as well as from useful feedback from Eva Ascarza and Peter Fader. 
\fi

\ifsubmission
\section*{Declaration of generative AI and AI-assisted technologies in the manuscript preparation process}
The first version of the working codebase underlying this paper was written entirely by the first author in late 2024. The authors subsequently used Anthropic's Claude Opus~3 and~4 to refactor this codebase and make targeted edits, manually reviewing all changes and confirming their correctness with unit tests. The authors further used Claude Fable~5 to implement plotting functions, to write code for submitting and managing experiments on a compute cluster, and to assist with exploratory data analysis in the two case studies, in all cases under close supervision of the authors. In preparing the manuscript itself, the authors used Claude Opus~3--4 and Claude Fable~5 to perform mechanical edits in \LaTeX{} (e.g., standardizing \textsc{BibTeX} entries and enforcing consistent notation and terminology across sections) and to proofread the text, catch typographical errors, and provide feedback. After using these tools, the authors reviewed and edited all content as needed and take full responsibility for the content of the published article.
\fi

\setlength{\bibsep}{0pt}
\bibliographystyle{abbrvnat}
\bibliography{references}

\ifsubmission
\else
\input{sections/appendix}
\fi
\end{document}

%% file: preamble.tex
\ifcsname ifsubmission\endcsname\else
  \expandafter\newif\csname ifsubmission\endcsname
\fi

\hfuzz=\maxdimen

\usepackage[utf8]{inputenc} 
\usepackage[T1]{fontenc}    

\ifsubmission
  \usepackage{amsmath}
  \usepackage{amsthm}   
  \usepackage{newtxtext}
  \usepackage{newtxmath}
  \usepackage[scaled=0.92]{PTSans}
\else
  \usepackage[bitstream-charter]{mathdesign}
  \usepackage{amsmath}
  \usepackage[scaled=0.92]{PTSans}
\fi

\usepackage[
  paper  = letterpaper,
  left   = 1.0in,
  right  = 1.0in,
  top    = 1.0in,
  bottom = 1.0in,
  ]{geometry}

\usepackage[usenames,dvipsnames,table]{xcolor}
\definecolor{shadecolor}{gray}{0.9}

\usepackage[final,expansion=alltext]{microtype}
\usepackage[english]{babel}
\usepackage[parfill]{parskip}
\usepackage{afterpage}
\usepackage{framed}
\usepackage{setspace}
\ifsubmission
  \setdisplayskipstretch{1}
  \everydisplay\expandafter{\the\everydisplay
    \linespread{1}\selectfont
    \ifdim\belowdisplayskip>6pt \belowdisplayskip=6pt plus 3pt minus 4pt\fi
    \ifdim\belowdisplayshortskip>2pt \belowdisplayshortskip=2pt plus 2pt minus 1pt\fi}
\fi

{\endMakeFramed}

\DeclareRobustCommand{\parhead}[1]{\textbf{#1}~}

\usepackage{lineno}

\usepackage{ragged2e}

\newcounter{parcount}

\usepackage{graphicx}
\usepackage{wrapfig}
\usepackage[labelfont=bf,font=footnotesize,width=.9\textwidth]{caption}
\usepackage[format=hang]{subcaption}

\usepackage{booktabs,multirow,multicol}       

\usepackage{nicefrac}
\usepackage{pifont}     
\usepackage{makecell}   
\usepackage{tcolorbox}
\tcbuselibrary{skins}
\definecolor{WTYDColor}{HTML}{1F77B4}  
\definecolor{PGGGColor}{HTML}{D62728}  
\newcommand{\yes}{\textcolor{OliveGreen}{\ding{51}}}
\newcommand{\no}{\textcolor{BrickRed}{\ding{55}}}

\newcommand{\colgloss}[1]{{\footnotesize\itshape\color{black!55}#1}}

\usepackage[algoruled,linesnumbered]{algorithm2e}
\usepackage{listings}
\usepackage{fancyvrb}
\fvset{fontsize=\normalsize}

\usepackage[colorlinks,linktoc=all]{hyperref}
\usepackage[all]{hypcap}
\hypersetup{citecolor=MidnightBlue}
\hypersetup{linkcolor=MidnightBlue}
\hypersetup{urlcolor=MidnightBlue}

\usepackage[nameinlink,capitalise]{cleveref}
\Crefname{section}{\S}{\S}
\crefname{AlgoLine}{line}{lines}
\Crefname{AlgoLine}{Line}{Lines}
\AtBeginDocument{}
\usepackage{etoolbox}
\makeatletter
\patchcmd{\nl}{\stepcounter{AlgoLine}\algocf@nl@sethref{\theAlgoLine}}{\refstepcounter{AlgoLine}\algocf@nl@sethref{\theAlgoLine}}{}{}
\patchcmd{\enl}{\stepcounter{AlgoLine}\algocf@nl@sethref{\theAlgoLine}}{\refstepcounter{AlgoLine}\algocf@nl@sethref{\theAlgoLine}}{}{}
\makeatother

\usepackage[acronym,nowarn]{glossaries}

\lstdefinestyle{mystyle}{
    commentstyle=\color{OliveGreen},
    keywordstyle=\color{BurntOrange},
    numberstyle=\tiny\color{black!60},
    stringstyle=\color{MidnightBlue},
    basicstyle=\ttfamily,
    breakatwhitespace=false,
    breaklines=true,
    captionpos=b,
    keepspaces=true,
    numbers=left,
    numbersep=5pt,
    showspaces=false,
    showstringspaces=false,
    showtabs=false,
    tabsize=2
}
\input{preamble/preamble_math}

\input{preamble/definitions_basic}

\ifcsname ifpreview\endcsname\else
  \expandafter\newif\csname ifpreview\endcsname
\fi
\ifsubmission
  \ifpreview
    \input{preamble/commenting}
  \else
    \input{preamble/nocommenting}

  \fi
\else
  \input{preamble/commenting}
\fi

%% file: preamble/preamble_math.tex
\usepackage{centernot}
\usepackage{amsthm}         
\usepackage{nicefrac}       
\usepackage{mathtools}      
\usepackage{amsbsy}         
\usepackage{amstext}        
\usepackage{thmtools}       
\usepackage{thm-restate}    

\begingroup
    \makeatletter
    \@for\theoremstyle:=definition,remark,plain\do{%
        \expandafter\g@addto@macro\csname th@\theoremstyle\endcsname{%
            \addtolength\thm@preskip\parskip
            }%
        }
\endgroup

\crefname{lemma}{lemma}{lemmas}
\Crefname{lemma}{Lemma}{Lemmas}
\crefname{thm}{theorem}{theorems}
\Crefname{thm}{Theorem}{Theorems}
\crefname{prop}{proposition}{propositions}
\Crefname{prop}{Proposition}{Propositions}
\crefname{assumption}{assumption}{assumptions}
\crefname{assumption}{Assumption}{Assumptions}

\usepackage{arydshln}
\makeatletter
\def\adl@drawiv#1#2#3{%
        \hskip.5\tabcolsep
        \xleaders#3{#2.5\@tempdimb #1{1}#2.5\@tempdimb}%
                #2\z@ plus1fil minus1fil\relax
        \hskip.5\tabcolsep}
\newcommand{\cdashlinelr}[1]{%
  \noalign{\vskip\aboverulesep
           \global\let\@dashdrawstore\adl@draw
           \global\let\adl@draw\adl@drawiv}
  \cdashline{#1}
  \noalign{\global\let\adl@draw\@dashdrawstore
           \vskip\belowrulesep}}
\makeatother

\renewcommand{\epsilon}{\varepsilon}

\declaretheorem[style=plain,numberwithin=section,name=Theorem]{theorem}

\declaretheorem[style=definition,sibling=theorem,name=Example]{example}

\newenvironment{example*}
 {\pushQED{\qed}\example}
 {\popQED\endexample}
\numberwithin{equation}{section}

%% file: preamble/definitions_basic.tex
\newcommand{\defeq}{\overset{\mathrm{def}}{=}}

\newcommand{\iidsim}{\overset{\mathrm{iid}}{\sim}}

\newcommand{\msdefeq}{\overset{\mathsmaller{\mathrm{def}}}{=}}

\DeclareMathOperator*{\argmin}{argmin}


%% file: preamble/commenting.tex
\definecolor{WowColor}{rgb}{.75,0,.75}
\definecolor{SubtleColor}{rgb}{0,0,.50}

\newcommand{\LATER}[1]{\textcolor{SubtleColor}{ {\tiny \bf ($\dagger$)} #1}}
\newcommand{\TBD}[1]{\textcolor{SubtleColor}{ {\tiny \bf (!)} #1}}
\newcommand{\PROBLEM}[1]{\textcolor{WowColor}{ {\bf (!!)} {\bf #1}}}

\newcounter{margincounter}
\newcommand{\displaycounter}{{\arabic{margincounter}}}
\newcommand{\incdisplaycounter}{{\stepcounter{margincounter}\arabic{margincounter}}}

\newcommand{\fTBD}[1]{\textcolor{SubtleColor}{$\,^{(\incdisplaycounter)}$}\marginpar{\tiny\textcolor{SubtleColor}{ {\tiny $(\displaycounter)$} #1}}}

\newcommand{\fPROBLEM}[1]{\textcolor{WowColor}{$\,^{((\incdisplaycounter))}$}\marginpar{\tiny\textcolor{WowColor}{ {\bf $\mathbf{((\displaycounter))}$} {\bf #1}}}}

\newcommand{\fLATER}[1]{\textcolor{SubtleColor}{$\,^{(\incdisplaycounter\dagger)}$}\marginpar{\tiny\textcolor{SubtleColor}{ {\tiny $(\displaycounter\dagger)$} #1}}}

%% file: preamble/nocommenting.tex
\newcommand{\LATER}[1]{\PackageError{nocommenting}{Leftover LATER comment}{Remove all LATER comments before final build}}
\newcommand{\fLATER}[1]{\PackageError{nocommenting}{Leftover fLATER comment}{Remove all fLATER comments before final build}}
\newcommand{\TBD}[1]{\PackageError{nocommenting}{Leftover TBD comment}{Remove all TBD comments before final build}}
\newcommand{\fTBD}[1]{\PackageError{nocommenting}{Leftover fTBD comment}{Remove all fTBD comments before final build}}
\newcommand{\PROBLEM}[1]{\PackageError{nocommenting}{Leftover PROBLEM comment}{Remove all PROBLEM comments before final build}}
\newcommand{\fPROBLEM}[1]{\PackageError{nocommenting}{Leftover fPROBLEM comment}{Remove all fPROBLEM comments before final build}}

%% file: macros.tex
\newcommand{\bz}{{\boldsymbol{z}}}

\newcommand{\bu}{{\boldsymbol{u}}}
\newcommand{\bg}{{\boldsymbol{g}}}
\newcommand{\bD}{{\boldsymbol{\mathcal{D}}}}
\newcommand{\btheta}{{\boldsymbol{\theta}}}
\newcommand{\bphi}{{\boldsymbol{\phi}}}
\newcommand{\bpsi}{{\boldsymbol{\psi}}}
\newcommand{\beps}{{\boldsymbol{\varepsilon}}}
\newcommand{\itt}{\Delta t}

\newcommand{\defeqinl}{\,\,\mathsmaller{\defeq}\,\,}
\newcommand{\iidsiminl}{\,\,\mathsmaller{\iidsim}\,\,}

\newcommand{\palive}{P(\textsc{alive})}

\newcommand{\msp}[1]{\mathsmaller{(#1)}}

\newif\iftrimcites
\trimcitesfalse   
\newcommand{\subcitep}[3][]{%
  \if\relax\detokenize{#1}\relax
    \iftrimcites\ifsubmission\citep{#2}\else\citep{#2,#3}\fi\else\citep{#2,#3}\fi
  \else
    \iftrimcites\ifsubmission\citep[#1][]{#2}\else\citep[#1][]{#2,#3}\fi\else\citep[#1][]{#2,#3}\fi
  \fi}

\usepackage[normalem]{ulem}

\SetKwComment{tcc}{\textcolor{black!55}{\#}~}{}
\SetKwComment{tcp}{\textcolor{black!55}{$\triangleright$}~}{}

\newlength{\cmtw}
\newcommand{\acom}[1]{\tcp*[r]{\makebox[\cmtw][l]{\normalsize #1}}}
\newcommand{\acomf}[1]{\tcp*[f]{\makebox[\cmtw][l]{\normalsize #1}}}

\SetKwFor{RepUntil}{repeat until}{}{}

\newtcolorbox{algobox}[2][WTYDColor]{%
  enhanced, colback=#1!5, colframe=#1!60!black, boxrule=0.8pt, arc=3pt,
  left=6pt, right=6pt, top=10pt, bottom=4pt, before skip=5pt, after skip=5pt,
  overlay unbroken and first={%
    \node[anchor=west, font=\scriptsize\sffamily\bfseries, text=#1!60!black,
          fill=#1!5, inner xsep=5pt, inner ysep=1.5pt]
      at ([xshift=10pt, yshift=-2.5pt]frame.north west) {#2};},
}

\newcommand{\optional}[2][black!55]{{\color{#1}#2}}

\ifcsname ifsubmission\endcsname\else
  \expandafter\newif\csname ifsubmission\endcsname
\fi
\newcommand{\apporweb}[2]{\ifsubmission #2\else\Cref{#1}\fi}

%% file: sections/0_abstract.tex
\begin{abstract}
Despite the wide variety of existing Buy-`Til-You-Die (BTYD) models, nearly all rely upon the convenient assumption of transactions following a Poisson process. As modern customer bases grow larger and more diverse, a major gap in the marketing literature is BTYD models that can account for heterogeneity in timing patterns across millions of customers. This paper addresses that gap, introducing a family of models that assume transactions follow a Weibull renewal process and developing a highly scalable scheme for parameter estimation based on an amortized variational inference procedure. The proposed model fits to a proprietary dataset of 5 million online retail customers in 8 minutes which would take the current state-of-the-art an estimated 3-4 days. We show both theoretically and empirically that this dramatic improvement in computational performance comes with no appreciable change to either model interpretation or predictive performance. Beyond scalability, gradient-based variational inference also makes it easy to extend the model to covariates, which we illustrate on a public dataset of 4 million political donors during the 2020 US General election cycle. More generally, this paper demonstrates how to blend recent advances in approximate Bayesian inference and the tools of modern machine learning to dramatically improve the efficiency and expressivity of probabilistic models for customer base analysis.~\looseness=-1 
\end{abstract}

\textbf{Keywords:} customer-base analysis; Buy `Til You Die models; regularity and clumpiness; Weibull renewal processes; amortized variational inference

%% file: sections/1_intro.tex
\section{Introduction}
\label{sec:intro}

Large-scale customer relationship management (CRM) data recording the behavior of millions or more customers is now commonplace in industries like banking~\citep{lemos2022propension}, telecommunications~\citep{verbeke2014social}, and many others. Although increasingly large in length, CRM data often remains highly sparse at the level of individual customers, due in part to the common pattern of attrition by which customers only actively transact for short periods of time, after which they churn or ``die''~\citep{reinartz2000profitability}. Large customer bases are moreover naturally heterogeneous, with many customers exhibiting a variety of different purchasing patterns~\citep{allenby1999marketing}. This combination of size, sparsity, and heterogeneity motivates statistical models of CRM data that strike a particular balance of being 1) \textit{scalable}, capable of being efficiently fit to millions of customers, 2) \textit{parsimonious}, allowing for statistical information to be pooled across sparsely observed units, but also 3) \textit{sufficiently expressive}, not collapsing salient and predictive heterogeneity in customer behavior.~\looseness=-1

Probabilistic models following the ``Buy `Til You Die'' (BTYD) framework of~\citet{schmittlein1987counting} are a mainstay of statistical marketing~\citep{gupta2006modeling,fader2009probability}, owing in part to their parsimonious representation of customer behavior. This representation splits cleanly into: 1) a transaction process, governing the timing by which a customer ``buys'', and 2) a lifetime process, governing the time until a customer ``dies''. A small set of such models---e.g., the original Pareto/NBD of \citeauthor{schmittlein1987counting}, or the BG/NBD model of~\citet{fader2005counting}---are simple enough to be fit scalably to large datasets using gradient-based optimization schemes which exploit such models' analytic tractability. While these ``workhorse'' models account for a large share of practical usage, a large literature has since introduced many new variants which tend to trade analytic and computational tractability for expressivity, introducing more flexible probabilistic assumptions that then necessitate approximate inference schemes like MCMC for parameter estimation~\citep[e.g.,][]{netzer2008hidden,abe2009counting,platzer2016ticking,dew2018bayesian}. 

Despite the wide variety of BTYD models to date however, very few depart from the analytically convenient assumption that the event times of an active customer's transactions follow a Poisson process. The Poisson has long been the canonical model for ``pure chance'' events in many fields, with departures from it indicating the presence of unaccounted for structure. In marketing, transaction patterns that are either more regular or more ``clumpy'' than expected under a Poisson process have been highlighted and studied for several decades~\citep[e.g.,][]{chatfield1973consumer,gupta1991stochastic,zhang2015predicting,platzer2016customer}. Nevertheless, the lack of scalable non-Poisson models for heterogeneous patterns of timing in large-scale customer bases remains a major gap.

This paper fills that gap by introducing a family of non-Poisson BTYD models which enjoy almost all the core benefits of the ``workhorse'' models---namely 1) they can be scalably fit to millions of customers, 2) their key quantities can be efficiently computed, and 3) they can be easily extended to covariates. In particular, this paper shows how to extend amortized variational inference~\citep{kingma2014auto,rezende2014stochastic,zhang2018advances} to such models, which turns the problem of posterior inference into a gradient-based optimization problem, toward which all the tools of modern machine learning---i.e., GPU accelerated computation, automatic differentiation, and neural networks---can be deployed.

This family of models is built around transactions following a Weibull renewal process, an assumption with precedent in the marketing literature~\citep{chatfield1973consumer,gupta1991stochastic,allenby1999dynamic,mcshane2008count} but one which departs from existing regularity-aware BTYD models which instead assume gamma-distributed renewals~\citep{platzer2016ticking,reutterer2021leveraging}. As we show, the shift from gamma to Weibull facilitates the development of gradient-based variational inference, and does so without any appreciable change to such models' expressivity, interpretation, or performance. We parameterize the Weibull in a nonstandard way so that the model emits a set of customer-specific latent variables with the same interpretation as in previous work, a correspondence which we characterize formally. In addition to the model, we also rely upon the Weibull distribution as the variational family during inference, where its convenient analytic properties contribute to a low-variance stochastic optimization procedure via pathwise or ``reparameterization'' gradient estimates~\citep{henderson2006handbooks,kingma2014auto}. Due to this dual reliance on the Weibull distribution, we dub the proposed approach ``Wei' `Til You Die'' (WTYD).~\looseness=-1

We illustrate the importance of modeling heterogeneous regularity in large-scale customer bases with two case studies on both proprietary and public data. The first case study is on 5 million customers of an online subscription-based retail firm, whose anonymized data we acquired from the marketing analytics firm Ocurate. The second is on 4 million individual political donors during the run-up to the 2020 US General elections, the data for which is publicly available through the Federal Elections Commission (FEC). In both cases, WTYD reveals interpretable heterogeneous structure in regularity, frequency, and churn that informs and accords with our knowledge of both domains. Fitting WTYD takes 8 minutes on the dataset of 5 million, which takes the current state-of-the-art an estimated 3-4 days. This comes with no appreciable loss in predictive performance, as we demonstrate in empirical comparisons on random subsets of the real data as well as on a suite of simulated datasets. On simulated data, where the ground-truth is known, we also show that WTYD exhibits no degradation in parameter or latent-variable recovery despite its dramatic improvement in computational efficiency. We also provide an illustration on how to extend WTYD to covariates and demonstrate the benefit in doing so.

\parhead{Roadmap.} The contributions of this paper are structured as follows.
\begin{itemize}
\item After defining notation and preliminaries on CRM data in~\Cref{sec:background_crm}, we give an overview of related BTYD models in \Cref{sec:background} which highlights a specific set of tradeoffs and connections.
\item We then introduce the WTYD model in \Cref{sec:weityd_dev} where we formally characterize its close correspondence to the Pareto/GGG model of~\citet{platzer2016customer} and give several key properties, including its likelihood and a principled approximation to expected lifetime value.
\item In \Cref{sec:inference}, we then develop an amortized variational inference procedure that fits WTYD by stochastic gradient-based optimization. This section develops the basic procedure based on parameterizing an inference network tailored to point process observations as well as on pathwise gradient estimators. It also introduces a series of algorithmic improvements which are instrumental to making the basic algorithm work in practice, including Rao-Blackwellization, which exploits analytic properties of the model to decrease gradient variance, and ``inference compilation''~\citep{le2017inference} which pretrains the inference network on synthetic data. Finally, this section introduces principled variational approximations to key posterior quantities such as ``P(alive)'', the probability a customer is alive given their transactions.
\item In \Cref{sec:performance_eval}, we report an extensive suite of simulation studies which compares WTYD to Pareto/GGG on recovering ground-truth parameters and latent-variables and forecasting future data across a range of data settings, showing that WTYD obtains a substantial speedup without any appreciable loss in performance.
\item In \Cref{sec:case_studies} we then provide two case studies on a novel proprietary dataset of 5 million online retail customers and a public dataset of 4 million political donors. In both, we show that WTYD recovers interpretable heterogeneous structure in regularity, frequency and churn, doing so highly efficiently. We also provide an example of how to extend WTYD to covariates.
\item Finally, we conclude in \Cref{sec:conclusion} by drawing further connections to the literature and outlining promising future avenues that build upon this work.
\end{itemize}

%% file: sections/2_data.tex
\begin{figure}[t!]
\includegraphics[width=\linewidth]{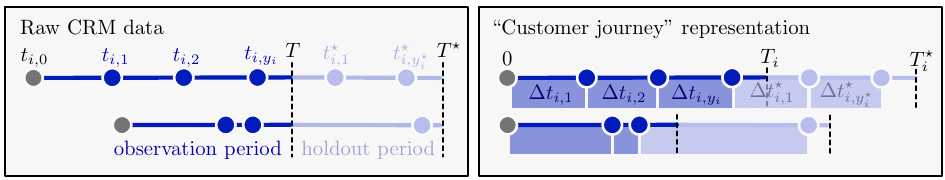}
\caption{Raw CRM data (left) based on absolute transaction timestamps versus the ``customer journey'' representation (right) based on inter-transaction times and varying total times.}
\label{fig:customer_journey}
\end{figure}

\section{Customer Relationship Management (CRM) Data}
\label{sec:background_crm}
CRM data consists of ``which customer bought what and when'' micro-records. As the timing pattern of such purchases is often the main object of statistical analysis, we will focus on the scenario where there is only one type of purchase to simplify notation and exposition.

A CRM database grows over time as new customers are added (or ``born'') and their transactions recorded. At any current time $T$, there are $n$ customers in the database, each of whom is indexed by $i \in [n]$ where $[n] \defeqinl \{1,\dots, n\}$. Let $t_{i,0} < T$ be the time that customer $i \in [n]$ is ``born'', often defined as the time of their first transaction. The times of their subsequent or ``repeat'' transactions are then $\mathbf{t}_i \defeqinl \{t_{i,1},\dots, t_{i,y_i}\}$ where $t_{i,j} < T$ for all $j \in [y_i]$. The time of their latest transaction $t_{i,y_i}$ is called their ``\textit{recency}'' while the total count of their repeat transactions during the observation window $y_i \in \mathbb{N}$ is called their ``\textit{frequency}''. Note that if $y_i = 0$ then $\mathbf{t}_i = \emptyset$.~\looseness=-1

For the purpose of evaluation and benchmarking of forecasting methods, we can also define a future holdout period up until time $T^\star > T$, such that the observation (or training) window is $(0, T)$ and the holdout (or testing) period is $(T, T^\star)$. We then denote the transaction times of customer $i$ during the holdout period to be $\mathbf{t}_i^\star \defeqinl \{t_{i,1}^\star, \dots t_{i,y_i^\star}^\star\}$ such that $t^\star_{i,j} \in (T, T^\star)$ for all $j \in [y_i^\star]$ and where $y_i^\star$ denotes their number of transactions during the holdout period, which we call the customer's ``\textit{future frequency}''. Note again that if $y^\star_i = 0$ then $\mathbf{t}_i^\star = \emptyset$.

Since customers are ``born'' into the data at different times, it is common to represent the timing of their transactions relativistically, which simplifies model specification. First, rather than deal in absolute timestamps, we consider the customer's \textit{inter-transaction times (ITTs)}, which for $y_i > 0$ in the observation window are denoted $\boldsymbol{\itt}_i \defeqinl (\itt_{i,1},\dots, \itt_{i,y_i})$ and defined as
\begin{equation}
\label{eq:itts}
\itt_{i,j} \defeq t_{i,j} - t_{i,\,j {-} 1} \,\,\textrm{ for }\,\, j \in [y_i] \,\,\textrm{ such that }\,\, t_{i,j} = t_{i,0} + \sum_{j'\leq j} \itt_{i,j'}.
\end{equation}
Similarly, the customer's ITTs during the holdout period are then 
$\boldsymbol{\itt}^\star_i \defeqinl (\itt^\star_{i,1},\dots, \itt^\star_{i,y^\star_i})$ for $y^\star_i > 0$, where the first of these is the gap between the customer's last transaction during the observation window and their first transaction during holdout---i.e., $\itt^\star_{i,1} \defeqinl t^\star_{i,1}-t_{i,y_i}$.

Considering only the ITTs is called the \textit{``customer's journey''} representation, which allows us to specify models that assume customers are ``born'' simultaneously at $t=0$. The length of each customer's observation window then varies, each of which is defined $T_i \defeqinl T-t_{i,0}$ and often called their \textit{``total time''}. By extension, $T_i^\star \defeqinl T^\star - t_{i,0}$ is their total time through the holdout period. For a visual representation of these quantities, see~\Cref{fig:customer_journey}.

%% file: sections/3_background.tex
\section{Buy `Til You Die (BTYD) Models}
\label{sec:background}
A common pattern of consumer behavior is \textit{churn} where customers eventually cease purchasing from a firm permanently. Accounting for this behavior is important---e.g., for valuing a customer base or even an entire firm~\subcitep{gupta2006modeling}{mccarthy2018customer}---especially in \textit{non-contractual} settings where the firm never directly observes a cancellation event~\citep{reinartz2000profitability}. Marketers have traditionally done so by modeling each customer as being active (or ``alive'') during a set interval $(t_{i,0}, \tau_i)$---i.e., between the time of their ``birth'' $t_{i,0}$ and ``death'' $\tau_i$. Critically, while a customer's birth is observed in CRM data, their death typically is not. Death manifests indirectly though, such as a long-churned customer having ``recency'' much less than the current time $t_{i,y_i} \ll T$. Thus, statistical models that treat $\tau_i$ explicitly as a latent variable are highly motivated.

A long history of such models in marketing, which collectively form the subfield of \textit{customer-base analysis}~\citep{fader2009probability}, follows the ``buy `til you die'' framework of~\citet{schmittlein1987counting}, wherein each model is fundamentally characterized by two processes:
\begin{enumerate}
\item a \textit{transaction process} for the timing of transactions $\mathbf{t}_i$ while customers are ``alive'', and
\item a \textit{lifetime (or dropout) process} for the time $\tau_i$ until customers ``die''.
\end{enumerate}

\textbf{Transaction renewal processes.} The vast majority of BTYD models assume transactions follow a Poisson process, or more generally a \textit{renewal process}~\subcitep{feller1971}{cox1962}, under which ITTs are drawn $\itt_{i,j} \,\mathsmaller{\iidsim}\, f_i$ from some renewal distribution $f_i$ specific to customer $i$. For instance, a Poisson transaction process is one where renewals are Exponential---i.e., $f_i \defeqinl \textrm{Exp}(\textsc{rate}\!=\!\lambda_i)$. This then induces a continuous-time \textit{renewal counting process} $Y_i(\cdot)$:
\begin{equation}
\label{eq:count_process}
\itt_{i,j} \iidsim f_i \,\,\,\textrm{ and }\,\,\, Y_i(t) \,\msdefeq\, \max\Big\{m : \sum_{j=1}^{m} \itt_{i,j} \leq t\Big\}.
\end{equation} 
The customer's ``frequency'' (as defined in the previous section) is then $y_i = Y_i(T_i \wedge \tau_i)$ where $a \wedge b \defeqinl \min\{a,b\}$. The driving assumption behind BTYD models---i.e., that a customer never buys again after $\tau_i$---is then expressed as the \textit{``customer's lifetime value'' (CLV)}\footnote{We adopt the convention of identifying CLV with the transaction count. The \emph{monetary} dimension of CLV (i.e., spend per transaction) is often modeled separately---e.g.\ via the Gamma--Gamma model~\citep{fader2013gamma}.} being $Y_i(\infty)=Y_i(\tau_i)$. We can also define the counting process on any interval $(s, t)$ for $s<t$ to denote $Y_i(s, t) \,\defeqinl  Y_i(t) - Y_i(s)$. For instance, a customer's ``future frequency'' is $y_i^\star = Y_i(T_i, T_i^\star)$.~\looseness=-1

\textbf{Key quantities.} Practitioners have traditionally sought to query BTYD models in a few key ways, which can be cleanly divided into \textit{prior vs posterior quantities}, the former being:
\begin{enumerate}[leftmargin=11em,label=(1\alph*)]
    \item Expected customer lifetime value: $\mathbb{E}[Y_i(\infty)]$, \label{item:clv}
    \item Marginal likelihood: $p(\mathcal{D}_i) = \int p(\mathcal{D}_i, \boldsymbol{z}_i) \,\textrm{d}\boldsymbol{z}_i$, \label{item:marglike}
\end{enumerate}
where $\mathcal{D}_i \defeqinl \{\boldsymbol{\itt}_i,\,T_i,\, \dots\}$ are the data on customer $i$ during the observation window, usually their ITTs and total time, but potentially other data (e.g., absolute timestamps $\mathbf{t}_i$, covariates $\mathbf{x}_i$), and $\boldsymbol{z}_i \defeqinl \{f_i, \tau_i,\, \dots\}$ are customer-specific latent variables, usually just their renewal distribution (or parameters thereof) and time of death. The key posterior quantities are then:
\begin{enumerate}[leftmargin=11em,label=(2\alph*)]
    \item Forecasted future frequency: $\mathbb{E}[Y_i(T_i, T_i^\star) \mid \mathcal{D}_i]$, \label{item:forecast}
    \item Posterior predictive distribution: $p(\mathcal{D}^\star_i \mid \mathcal{D}_i) = \int p(\mathcal{D}^\star_i, \boldsymbol{z}_i \mid \mathcal{D}_i)  \,\textrm{d}\boldsymbol{z}_i$, \label{item:postpred}
    \item Probability that customer $i$ is alive ``$\textrm{P} (\textsc{alive})$'': $p(\tau_i > T_i \mid \mathcal{D}_i)$, \label{item:palive}
\end{enumerate}
where $\mathcal{D}^\star_i \defeqinl \{\boldsymbol{\itt}^\star_i,\,T^\star_i,\, \dots\}$ are the data on customer $i$ during the holdout period. 

Being able to efficiently compute these quantities facilitates the use of such models. The expected frequencies in \ref{item:clv} and \ref{item:forecast} are interpretable summary statistics used by marketers to segment the customer base and make predictions, with expected CLV in \ref{item:clv} being particularly useful when customer-level covariates are available to help overcome the ``cold-start'' prediction problem. Tractable computation of the prior and posterior predictive distributions in \ref{item:marglike} and \ref{item:postpred} then facilitates parameter estimation and model selection, as it does with most probabilistic models. Finally, \palive~in~\ref{item:palive} plays a particularly central role in BTYD models, as computing it is a subroutine for computing most other key quantities, and it is itself an interpretable and useful summary statistic useful.~\looseness=-1

Whether the quantities above are tractable depends on the probabilistic assumptions of a given model. In the rest of this section, we survey some well-known BTYD models, emphasizing those we build upon, and highlighting the tension between tractability and expressivity that has been a key theme in this literature and which this paper seeks to ameliorate.

\textbf{Pareto/NBD.} The original BTYD model of~\citet{schmittlein1987counting} assumed that transactions follow a Poisson process until a Pareto-distributed time-of-death---i.e.,
\begin{modelbox}[gray]{Lifetime process}{Pareto/NBD}
\begin{align}
\label{eq:pareto}
\tau_i &\sim \textrm{Pareto}(\textsc{shape}\!=\!s^{\msp{\tau}}, \textsc{scale}\!=\!r^{\msp{\tau}})
\end{align}
\end{modelbox}
\begin{modelbox}[gray]{Transaction process}{Pareto/NBD}
\begin{align}
\label{eq:exprenewals}
\itt_{i,j} &\iidsim \textrm{Gamma}(\textsc{shape}\!=\!1,\,\textsc{rate}\!=\!\lambda_i)
\end{align}
\end{modelbox}
where we note that a Gamma distribution with shape equal to one is equivalent to an Exponential. Conditional on the customer-specific latent variables---i.e., $\boldsymbol{z}_i \defeqinl \{\tau_i, \lambda_i\}$---this induces a Poisson distribution over the frequency---i.e., $y_i \sim \textrm{Pois}(\textsc{mean}=\lambda_i(\tau_i \wedge T_i))$. The model then further posits a conjugate Gamma prior over the heterogeneous rates:
\begin{modelbox}[gray]{Heterogeneity prior}{Pareto/NBD}
\begin{align}
\label{eq:lambdaprior}
\lambda_i &\sim \textrm{Gamma}(\textsc{shape}\!=\!s^{\msp{\lambda}},\, \textsc{rate}\!=\!r^{\msp{\lambda}})
\end{align}
\end{modelbox}
Marginalizing out $\lambda_i$ then induces a \underline{N}egative \underline{B}inomial \underline{d}istribution over $y_i$, and the model is thus named ``Pareto/NBD''. The set of model parameters is $\btheta \defeqinl \{s^{\msp{\tau}}, r^{\msp{\tau}}, s^{\msp{\lambda}}, r^{\msp{\lambda}}\}$.

All of the key expressions listed above are analytic. Parameter estimation can thus proceed by directly maximizing the marginal likelihood via gradient ascent\footnote{Strictly speaking, the positive model parameters are optimized on an unconstrained scale via log transformation.}---e.g.,
\begin{equation}
\btheta \leftarrow {\textsc{Update}\Big(\btheta,\,\, \textstyle\sum_{i=1}^n \nabla_\btheta \log p_\btheta(\mathcal{D}_i)\Big)}. 
\end{equation}
Here $\textsc{Update}(\boldsymbol{w}, \boldsymbol{g})$ denotes any single step of a first-order optimizer that moves the parameters $\boldsymbol{w}$ along the gradient $\boldsymbol{g}$ of the training objective (e.g., vanilla gradient ascent, Adam).  This is straightforward to implement and scale to large datasets using modern frameworks. 

\textbf{Incorporating covariates.} Another benefit of gradient-based parameter estimation is that it makes the model particularly easy to modify so that its parameters are functions of customer-level covariates $\btheta_i \defeqinl \btheta(\boldsymbol{x}_i)$. For instance, \citet{fader2007incorporating} recommend setting
\begin{equation}
\label{eq:covariates}
\btheta_i = 
\begin{bmatrix*}[l]
           s_i^{\msp{\tau}} \\[0.25em]
           r_i^{\msp{\tau}} \\[0.25em]
           s_i^{\msp{\lambda}} \\[0.25em]
           r_i^{\msp{\lambda}}
\end{bmatrix*}
= \btheta(\boldsymbol{x}_i) =
\begin{bmatrix*}[l]
    s_0^{\msp{\tau}} \\[0.25em]
    r_0^{\msp{\tau}} \exp(-\boldsymbol{x}_i^\top \boldsymbol{\beta}^{\msp{\tau}}) \\[0.25em]
    s_0^{\msp{\lambda}} \\[0.25em]
    r_0^{\msp{\lambda}} \exp(-\boldsymbol{x}_i^\top \boldsymbol{\beta}^{\msp{\lambda}})
\end{bmatrix*},
\end{equation}
where the learnable parameters in $\btheta(\cdot)$ are the weights $\{\boldsymbol{\beta}^{\msp{\tau}}, \boldsymbol{\beta}^{\msp{\lambda}}\}$ and biases $\{s_0^{\msp{\tau}}, r_0^{\msp{\tau}}, s_0^{\msp{\lambda}}, r_0^{\msp{\lambda}}\}$. Many other settings are possible---e.g., $\btheta(\cdot)$ could be a feedforward neural network---and can be fit via gradient ascent as above provided that $\btheta(\cdot)$ defines a differentiable map.

\textbf{Pareto/NBD ``the easy way'' and related Poisson models.} While the Pareto/NBD model is entirely analytic, the derivation and implementation of its key expressions can be ``rather daunting''~\citep{fader2005note}. \citet{fader2005counting} thus introduced the BG/NBD model---dubbing it ``Pareto/NBD the easy way''---which replaces the lifetime process in~\cref{eq:pareto} by assuming the death time coincides exactly with a transaction time $\tau_i \in \{t_{i,1}, t_{i,2},\dots\}$, with which one being a latent variable drawn according to a \underline{B}eta-mixture of \underline{G}eometric distributions---i.e.,~\looseness=-1 
\begin{modelbox}[gray]{Lifetime process}{(M)BG/NBD}
\begin{equation}
\label{eq:bg}
J_i \sim \textrm{BetaGeometric}(a_\tau, b_\tau) \textrm{ and }\tau_i = t_{i, \,J_i}
\end{equation}
\end{modelbox}
Doing this greatly simplifies the model. However, it does so at the expense of tying death time to the transaction times. A simple \underline{m}odification called MBG/NBD~\citep{batislam2007empirical} further allows death to occur before any transaction. In both cases however, by tying the death and transaction times, the (M)BG/NBD model assumes high-volume customers churn faster~\citep{jerath2011new}, an assumption not made by Pareto/NBD and one that is often violated in practice.~\looseness=-1

Due to their analytic tractability, the Pareto/NBD and (M)BG/NBD models have long been the ``workhorses'' of CRM modeling, with mature implementations in both 
R~\citep[e.g.,][]{dziurzynski2014btyd,platzer2016customer} and Python~\subcitep[e.g.,]{lifetimes}{deanpymc}. A key benefit of these models is that they emit a sufficient statistic representation such that the data on each customer need only consist of their frequency, recency, and total time---i.e., $\mathcal{D}_i \defeqinl \{y_i, t_{i,y_i}, T_i\}$. This further improves these models' scalability and ease, allowing even for implementations directly in Excel~\citep{fader2005implementing}.~\looseness=-1

Many extensions of the ``workhorse'' models have since been proposed---e.g., hierarchical Bayesian variants with covariates~\citep{abe2009counting}, models with time-varying covariates~\citep{schweidel2013incorporating}, or Bayesian nonparametric models~\citep{dew2018bayesian}---see surveys by \citet{fader2009probability} and \citet{casteran2021modeling}, among others. A key theme in this literature is that making the ``workhorse'' models more expressive comes at the cost of tractability and scalability. In particular, the assumption of a Poisson transaction process is key to numerous analytic properties in these models and is thus rarely relaxed.

\textbf{Non-Poisson transactions.} As has been observed for several decades, many real-world transaction patterns are poorly modeled by a Poisson process, exhibiting more or less regularity in their timing than would be expected by Exponentially-distributed renewals~\citep{chatfield1973consumer,gupta1991stochastic,allenby1999dynamic}. On one side, subscription-based customers transact with great regularity, exhibiting an ITT distribution that is highly peaked around its mean. On the other side, ``binge watching'' customers transact many times in short succession followed by long periods of inactivity~\citep{schweidel2016binge}, exhibiting a pattern of ITTs known as ``clumpiness''~\citep{zhang2015predicting}.~\looseness=-1

\textbf{Pareto/GGG.} Writing the Poisson process assumption as we do in~\cref{eq:exprenewals}, it is straightforward to modify it to account for regularity by introducing one extra customer-specific latent variable $k_i$ and defining the ITTs to then be drawn from a fully flexible Gamma distribution as:
\begin{modelbox}[PGGGColor]{Transaction process}{Pareto/GGG}
\begin{align}
    \label{eq:pggg}
\itt_{i,j} &\iidsim \textrm{Gamma}(\textsc{shape}\!=\!k_i,\,\textsc{rate}\!=\!k_i\lambda_i) 
\end{align}
\end{modelbox}
While the mean remains unchanged $\mathbb{E}[\itt_{i,j}]=\nicefrac{1}{\lambda_i}$, the concentration of the ITT distribution around its mean is now controlled by $k_i$, with $k_i>1$ and $k_i<1$ yielding more regular versus ``clumpy'' timing patterns, respectively. This model was introduced by~\citet{platzer2016ticking}, who additionally posit the following prior over the customer's ``regularity'':
\begin{modelbox}[PGGGColor]{Heterogeneity prior}{Pareto/GGG}
\begin{align}
k_i &\sim \textrm{Gamma}(\textsc{shape} \!=\!s^{\msp{k}},\,\textsc{rate}\!=\!r^{\msp{k}})
\end{align}
\end{modelbox}
The model is named ``Pareto/GGG'' as it retains the Pareto lifetime process and \underline{G}amma prior over $\lambda_i$ while adding a \underline{G}amma prior over $k_i$ and making the renewal distribution \underline{G}amma. The model retains a sufficient statistic representation, requiring only one extra ``regularity'' statistic---i.e., $\mathcal{D}_i \defeqinl \{y_i, t_{i,y_i}, \kappa_i, T_i\}$---where $\kappa_i \defeqinl \sum_j \log \itt_{i,j}$ is the sum-of-log-ITTs.

While conceptually simple, the move to a full Gamma renewal process makes all its key expressions non-analytic. In particular, its marginal likelihood $p_\btheta(\mathcal{D}_i)$ is intractable and thus estimation of its model parameters $\btheta \defeqinl \{s^{\msp{\tau}}, r^{\msp{\tau}}, s^{\msp{\lambda}}, r^{\msp{\lambda}}, s^{\msp{k}}, r^{\msp{k}}\}$ cannot proceed via direct gradient ascent, as it does with Pareto/NBD. Instead, \citeauthor{platzer2016ticking} provide an MCMC procedure that returns samples of $\btheta$ alongside samples of the customer-specific latent variables $\boldsymbol{z}_i \defeqinl \{\tau_i, \lambda_i, k_i\}$. The move to MCMC makes parameter estimation substantially more computationally intensive and less scalable. It also makes it harder to incorporate covariates into the model, which is not an available option in \textsc{BTYDplus}~\citep{platzer2016customer}, the main open-source implementation. 

\textbf{Pareto/GGG ``the easy way''.} In an effort to recover analytic tractability, \citet{reutterer2021leveraging} introduced an ``easy way'' in which the lifetime process is changed to match the (M)BG/NBD model in~\cref{eq:bg} and the transaction process is modified so that 1) all customers have the same regularity $k_i=k$ where $k$ is treated as a global model parameter, and 2) $k$ is constrained to only take positive integer values $k \in \{1,2,\dots\}$, thus inducing an Erlang-$k$ renewal distribution, a special case of the Gamma. The model is analytic in more (though still not all) of the key expressions, most notably the marginal likelihood $p_\btheta(\mathcal{D}_i)$ which can now be optimized directly. However, it is worth noting that the latent variables being marginalized out $\boldsymbol{z}_i \defeqinl \{\tau_i, \lambda_i\}$ no longer include customer-specific regularities $k_i$ and that the global regularity is constrained $k \geq 1$ so that it cannot express ``clumpy'' timing patterns. Moreover, the resulting model---named the (M)BG/CNBD-$k$ for the \underline{C}ondensed \underline{N}egative \underline{B}inomial-\underline{$k$} frequency distribution~\citep{schmittlein1983prediction} that the Erlang-$k$ renewals induce---suffers the same drawbacks as the (M)BG/NBD regarding the dependence between the lifetime and transaction processes.

\begin{table}[t]
\centering
\footnotesize
\setlength{\tabcolsep}{3.5pt}
\renewcommand{\arraystretch}{1.1}
\begin{tabular}{@{}l r r r r r@{}}
\toprule
\makecell[tl]{}
 & \makecell{1) Renewals $f_i$\\ \colgloss{\yes\;non-Poisson}}
 & \makecell{2) Data $\mathcal{D}_i$\\ \colgloss{\yes\;sufficiency}}
 & \makecell{3) Latents $\boldsymbol{z}_i$\\ \colgloss{\yes\;regularity}}
 & \makecell{4) Parameters $\btheta$\\ \colgloss{\yes\;covariates $\btheta(\boldsymbol{x}_i)$}}
 & \makecell{5) Inference\\ \colgloss{\yes\;scalable}} \\
\midrule
\addlinespace[1pt]
\makecell[tl]{Pareto/NBD\\[-0.6em]\colgloss{\tiny \citet{schmittlein1987counting}}}
 & Exponential\,\no
 & $\{y_i, t_{i,y_i}, T_i\}$\,\yes
 & $\{\tau_i, \lambda_i\}$\,\no
 & $\{s^{\msp{\tau}}, r^{\msp{\tau}}, s^{\msp{\lambda}}, r^{\msp{\lambda}}\}$\,\yes
 & MLE\,\yes \\
\makecell[tl]{(M)BG/NBD\\[-0.6em]\colgloss{\tiny \citet{fader2005counting}}}
 & Exponential\,\no
 & $\{y_i, t_{i,y_i}, T_i\}$\,\yes
 & $\{\tau_i, \lambda_i\}$\,\no
 & $\{a_\tau, b_\tau, s^{\msp{\lambda}}, r^{\msp{\lambda}}\}$\,\yes
 & MLE\,\yes \\
\makecell[tl]{Pareto/GGG\\[-0.6em]\colgloss{\tiny \citet{platzer2016ticking}}}
 & Gamma\,\yes
 & $\{\kappa_i, y_i, t_{i,y_i}, T_i\}$\,\yes
 & $\{\tau_i, \lambda_i, k_i\}$\,\yes
 & $\{s^{\msp{\tau}}, r^{\msp{\tau}}, s^{\msp{\lambda}}, r^{\msp{\lambda}}, s^{\msp{k}}, r^{\msp{k}}\}$\,\no
 & MCMC\,\no \\
\makecell[tl]{(M)BG/CNBD-$k$\\[-0.6em]\colgloss{\tiny \citet{reutterer2021leveraging}}}
 & Erlang-$k$\,\yes
 & $\{\kappa_i, y_i, t_{i,y_i}, T_i\}$\,\yes
 & $\{\tau_i, \lambda_i\}$\,\no
 & $\{s^{\msp{\tau}}, r^{\msp{\tau}}, s^{\msp{\lambda}}, r^{\msp{\lambda}}, k\}$\,\no
 & MLE\,\yes \\
\makecell[tl]{CLVAE\\[-0.6em]\colgloss{\tiny \citet{naf2026clvae}}}
 & Exponential\,\no
 & $\{y_i, t_{i,y_i}, T_i\}$\,\yes
 & $\{\tau_i, \lambda_i\}$\,\no
 & $\{s^{\msp{\tau}}, r^{\msp{\tau}}, s^{\msp{\lambda}}, r^{\msp{\lambda}}\}$\,\yes
 & VI\,\yes \\
\makecell[tl]{\textbf{WTYD} (this paper)} 
 & Weibull\,\yes
 & $\{\boldsymbol{\itt}_i, T_i\}$\,\no
 & $\{\tau_i, \lambda_i, k_i\}$\,\yes
 & $\{s^{\msp{\tau}}, r^{\msp{\tau}}, s^{\msp{\lambda}}, r^{\msp{\lambda}}, s^{\msp{k}}, r^{\msp{k}}\}$\,\yes
 & VI\,\yes \\
\bottomrule
\end{tabular}
\caption{BTYD models trade off different aspects of tractability and expressivity. \textbf{From left to right, models do~\yes~or do not~\no}: 1) assume more expressive transaction processes than Poisson, 2) yield sufficient statistics, 3) express heterogeneous regularity, 4) make it easy to incorporate  covariates, and 5) adopt scalable inference methods that are easily adapted to stream over minibatches of data.}
\label{tab:btyd_comparison}
\end{table}

\textbf{Summary of tradeoffs.} A key theme in the BTYD literature has been the tradeoff between analytic tractability and model expressivity. The original ``workhorse'' models are highly analytic, emitting sufficient statistics, scalable maximum likelihood-based inference, and allowing for the incorporation of covariates. However, these models and most of their extensions are premised on the key assumption that transactions follow a Poisson process. Models can achieve better fit when relaxing this assumption, but inevitably do so at some cost. \Cref{tab:btyd_comparison} summarizes these tradeoffs. Pareto/GGG generalizes transaction renewals to follow fully flexible Gamma distributions with heterogeneous regularity and does so while retaining sufficiency. However, the model requires MCMC to fit, which is not scalable and hampers incorporation of covariates. (M)BG/CNBD-$k$ meanwhile retains sufficiency while recovering a scalable inference scheme, but does so at the cost of constraining regularity to be homogeneous $k_i=k$ and non-``clumpy'' $k>1$. In what follows, we introduce a model---Wei' `Til You Die (WTYD)---that serves as a drop-in replacement for Pareto/GGG, adopting Weibull rather than Gamma renewals. This small change enables a form of scalable gradient-based variational inference (VI) for parameter estimation. The only cost is then to sufficiency: fitting WTYD requires conditioning on all ITTs $\boldsymbol{\itt}_i$. However, we argue that this tradeoff is highly advantageous over the alternatives, in light of modern datasets with large $n$ and modern computational infrastructure with low memory costs.~\looseness=-1

%% file: sections/4_model.tex
\section{Wei' `Til You Die (WTYD)}
\label{sec:weityd_dev}
The model we consider assumes that a customer's inter-transaction times (ITTs) are drawn
\begin{modelbox}[WTYDColor]{Transaction process}{WTYD}
\begin{align}
\label{eq:wtyditt}
\itt_{i,j} \iidsim \textrm{Weibull}\big(\textsc{shape} \!=\! k_i,\,\textsc{rate} \!=\! \lambda_i\,\Gamma(1+\nicefrac{1}{k_i})\big)
\end{align}
\end{modelbox}
and thus follow a Weibull renewal process~\subcitep{weirenewanalysis1964}{weirenewprocesses1994}. The Weibull distribution~\citep{weibull1951statistical} has density function $\textrm{Weibull}(w;\, \textsc{shape}\!=\!s,\, \textsc{rate}\!=\!r) = s\, r^s\, w^{s-1}\, \exp(-r^sw^s)$ and expected value $\mathbb{E}[w \mid s, r] = \Gamma(1+\nicefrac{1}{s})\,r^{-1}$ where $\Gamma(\cdot)$ is the gamma function.

Although it has not received much traction, the idea to model transaction processes with Weibull renewals is an old idea dating back at least to~\citet{allenby1999dynamic}, who use a Generalized Gamma renewal process (which generalizes the Weibull), or to~\citet{mcshane2008count} who model transaction frequencies with the Weibull Count distribution which arises from Weibull renewals. The definition of shape and rate in~\cref{eq:wtyditt}---where both are a function of $k_i$---departs from previous work, but will yield an appealing correspondence which we detail below.

We then posit the same lifetime process and priors as Pareto/GGG, repeated here for ease:
\begin{modelbox}[WTYDColor]{Lifetime process}{WTYD}
\begin{align}
\label{eq:wtyddeath}
\tau_i \sim \textrm{Pareto}(\textsc{shape}=s^{\msp{\tau}}, \textsc{scale}=r^{\msp{\tau}})
\end{align}
\end{modelbox}
\begin{modelbox}[WTYDColor]{Heterogeneity Prior}{WTYD}
\begin{align}
\label{eq:wtyd_kprior}
k_i &\sim \textrm{Gamma}(\textsc{shape}=s^{\msp{k}},\, \textsc{rate}=r^{\msp{k}})\\ 
\label{eq:wtyd_lamprior}
\lambda_i &\sim \textrm{Gamma}(\textsc{shape}=s^{\msp{\lambda}},\, \textsc{rate}=r^{\msp{\lambda}})
\end{align}
\end{modelbox}

\begin{figure}[t]
\includegraphics[width=\linewidth]{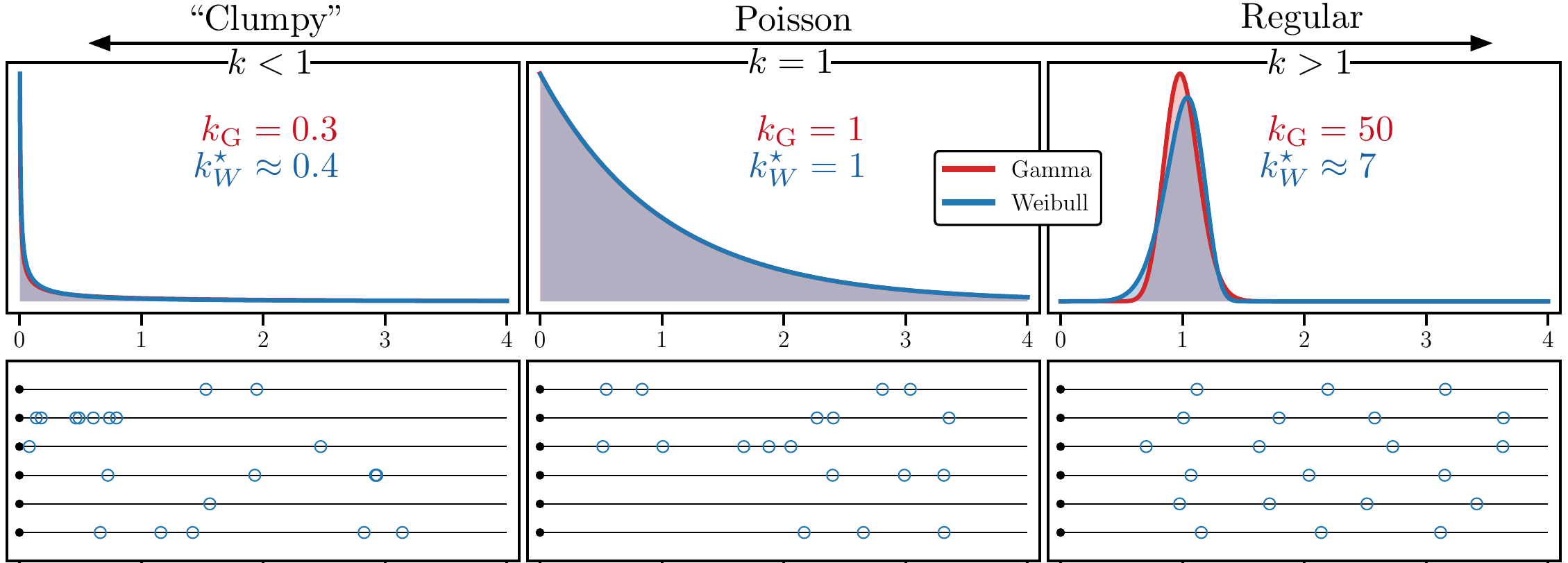}
\caption{Weibull and Gamma renewal models are very similar, with the best-fit Weibull to any given Gamma distribution having the same mean $\nicefrac{1}{\lambda_i}$ and concentration $k_i$ on the same side of 1. \textbf{Top:} Examples of a Gamma and best-fit Weibull for the three conditions on $k_i$, all with $\lambda_i=1$. \textbf{Bottom:} Six independent draws from the corresponding Gamma renewal process to illustrate the three conditions.}
\label{fig:weibull_renewalprocess}
\end{figure}

\textbf{Correspondence to Pareto/GGG.} The only difference between WTYD and Pareto/GGG is the Weibull versus Gamma transaction process. 
The Weibull and Gamma distributions are very similar, to the point of being difficult to distinguish from data \subcitep{bain1980probability}{fearn1991maximum}. Moreover, our parameterization of the Weibull in~\cref{eq:wtyditt} matches that of the Gamma in~\cref{eq:pggg} in terms of a \emph{mean--concentration} parameterization---i.e.,
\begin{align*}
\text{Weibull}\big(\textsc{shape}\!=\!k_i,\, \textsc{rate}\!=\!\lambda_i\Gamma(1+\nicefrac{1}{k_i})\big) &\,\,\Longleftrightarrow\,\, \text{Weibull}(\textsc{mean}\!=\!\nicefrac{1}{\lambda_i},\, \textsc{concentration}\!=\!k_i), \\[0.5em]
\text{Gamma}\big(\textsc{shape}\!=\!k_i,\, \textsc{rate}\!=\!k_i\lambda_i\big) &\,\,\Longleftrightarrow\,\, \text{Gamma}(\textsc{mean}\!=\!\nicefrac{1}{\lambda_i},\, \textsc{concentration}\!=\!k_i),
\end{align*}
where $\nicefrac{1}{\lambda_i}$ is the mean of customer $i$'s ITTs and $k_i$ is their ``concentration'', a term we use to mean any measure of spread which is independent of the mean. For the Gamma in~\cref{eq:pggg} the squared coefficient of variation (CV) is $\nicefrac{1}{k_i}$ while for the Weibull in~\cref{eq:wtyditt} it is $\tfrac{\Gamma(1+\nicefrac{2}{k_i})}{\Gamma(1+\nicefrac{1}{k_i})^2} -1$, neither of which depends on $\lambda_i$.

Both models therefore have the same set of latents $\boldsymbol{z}_i \defeqinl \{\tau_i, \lambda_i, k_i\}$, with each playing the same qualitative role. This correspondence can be sharpened further by noting that the concentration parameter $k_i$ acts as an \textit{index of regularity} in the same way for both models, yielding ``clumpy'' renewals for $k_i<1$, regular for $k_i > 1$, and a Poisson process at $k_i=1$. Moreover, we can show that the best-fit Weibull to any given Gamma, in terms of KL divergence, has equal mean and a concentration on the same side of 1, with a monotone mapping between them.~\looseness=-1

\begin{prop}[Correspondence of regularity index]
\label{prop:correspondance_prop}
Let $\nicefrac{1}{\lambda_W^\star}$ and $k_W^\star$ be the mean and concentration of a Weibull which minimizes the KL divergence to a Gamma with mean $\nicefrac{1}{\lambda_G}$ and concentration $k_G$---i.e., $
k_W^\star, \lambda_W^\star \,\,\in\,\, \argmin_{k_W, \lambda_W} \,\,\textit{KL}\Big(\textrm{Weibull}(\nicefrac{1}{\lambda_W}, k_W) \,\big\lvert \big\lvert\,  \textrm{Gamma}(\nicefrac{1}{\lambda_G}, k_G)\Big)$.
The minimizer is unique, and defined below where $\psi(\cdot)$ is the digamma function and $\gamma$ is Euler's constant: 
\begin{equation}
    \label{eq:kmap}
\lambda_G = \lambda_W^\star  \,\,\textrm{ and } \,\, k_G = k_W^\star \,\,\big(\gamma + \psi(1 + \nicefrac{1}{k_W^\star})\big)^{-1},
\end{equation}
Moreover the map $k_G \mapsto k_W^\star$ is strictly increasing, with equality at 1, and therefore:
\begin{align*}
&\textrm{(``clumpy'')} & &\textrm{(random)}& &\textrm{(regular)}\\[-0.25em]
k_W^\star < 1 &\hspace{0.75em}\Longleftrightarrow\hspace{0.5em} k_G < 1 & k_W^\star = 1 &\hspace{0.5em}\Longleftrightarrow\hspace{0.5em} k_G = 1 & k_W^\star > 1 &\hspace{0.5em}\Longleftrightarrow\hspace{0.5em} k_G > 1.
\end{align*}
\end{prop}
See~\apporweb{subsec:app_model}{Supplementary Material~A.1} for the proof, with the takeaway being that WTYD provides a drop-in replacement for Pareto/GGG with latent variables carrying the same qualitative meaning.

\textbf{Individual-level likelihood.} The likelihood of the model can be written:
\begin{equation}
\label{eq:indivlikelihood}
p(\mathcal{D}_i \mid \bz_i)
= \exp\!\Big(-r_i^{k_i}\,\big[(T_i \wedge \tau_i) - t_{i,y_i}\big]^{k_i}\Big)
\left[\frac{k_i^{y_i}\, r_i^{y_ik_i}\, \prod_{j=1}^{y_i} \itt_{i,j}^{k_i-1}}{\exp\!\big(r_i^{k_i}\sum_{j=1}^{y_i}\itt_{i,j}^{k_i}\big)}\right]^{\mathbb{1}(y_i > 0)}
\end{equation}
where we have aliased the rate parameter $r_i \msdefeq \lambda_i \Gamma(1+\nicefrac{1}{k_i})$. Like the Pareto/GGG, the marginal likelihood $p_\theta(\mathcal{D}_i) = \int p(\mathcal{D}_i \mid \bz_i) \, p_\theta(\boldsymbol{z}_i)\,\textrm{d}\boldsymbol{z}_i$ is intractable, but can be evaluated approximately via Monte Carlo. Unlike the Pareto/GGG however, the likelihood in~\cref{eq:indivlikelihood} involves only elementary functions beyond the gamma function, which every major automatic differentiation framework supports---a fact which will become key in developing scalable inference. 

\textbf{Loss of sufficiency.} Note that the likelihood does not emit a sufficient statistic representation and must input the customer's variable-length set of ITTs $\mathcal{D}_i \defeqinl \{\boldsymbol{\itt}_i, T_i\}$. This is due to the fact that the two-parameter Weibull is not an exponential family distribution. By contrast, the Pareto/GGG model---whose Gamma renewal distribution is an exponential family---conditions only on a fixed-length input $\mathcal{D}_i \defeqinl \{y_i, t_{y_i}, \kappa_i, T_i\}$. Thus, although the two models are equivalent in expressivity, WTYD incurs this disadvantage, which is practically important when customers have long transaction sequences. Nevertheless, we will show that this loss in sufficiency is made up for by the model becoming amenable to a form of gradient-based inference that is highly scalable.

\textbf{Expected CLV.} The expected customer lifetime value is also intractable, as with Pareto/GGG. However, writing it as $\mathbb{E}_\btheta\left[Y_i(\infty)\right] = \mathbb{E}_{\bz_i \sim p_\btheta(\bz_i)}[\mathbb{E}\left[Y_i(\tau_i) \mid \bz_i\right]]$, the inner expectation is the expectation of a Weibull Count random variable which can be evaluated via a recurrence given by~\citet{mcshane2008count}. Thus, the Monte Carlo estimator for CLV given below can be Rao-Blackwellized by sampling $\bz_i^{(s)} \iidsiminl p_\theta(\bz_i)$ and evaluating the inner expectation:
\begin{equation}
\label{eq:clv_mc}
\widehat{\mathbb{E}}^{\textrm{MC}}_\btheta\!\left[Y_i(\infty)\right]
\;\defeq\; \frac{1}{S}\sum_{s=1}^S \mathbb{E}\!\left[Y_i(\tau_i^{(s)}) \,\big|\, \bz_i^{(s)}\right].
\end{equation}
Alternatively, for large $\tau_i$, we can apply an elementary renewal theorem~\citep{feller1971} to derive an approximation that is a simple closed-form function of model parameters:
\begin{equation}
\label{eq:clv_rt}
\widehat{\mathbb{E}}^{\textrm{RT}}_\btheta\left[Y_i(\infty)\right]
\;\defeq\; \mathbb{E}_{\btheta}[\tau_i\lambda_i]
\;=\; \tfrac{r^{\msp{\tau}}}{s^{\msp{\tau}} - 1}\tfrac{s^{\msp{\lambda}}}{r^{\msp{\lambda}}}.
\end{equation}
Moreover, applying another standard renewal theorem~\citep{lorden1970excess}, we can derive a simple bound on the error of this estimator, which is particularly small for non-``clumpy'' populations:
\vspace{-\baselineskip}
\begin{prop}[{Error of $\widehat{\mathbb{E}}^{\textrm{RT}}_\btheta[Y_i(\infty)]$}]
\label{prop:clv_bias}
The error of the renewal-theoretic estimator is bounded by
\begin{equation}
\label{eq:clv_bias_bound}
\big|\widehat{\mathbb{E}}^{\textrm{RT}}_\btheta\!\left[Y_i(\infty)\right] - \mathbb{E}_\btheta\!\left[Y_i(\infty)\right]\big|
\;\leq\; \mathbb{E}_{k_i \sim p_\btheta(k_i)}\!\left[\frac{\Gamma(1+\nicefrac{2}{k_i})}{\Gamma(1+\nicefrac{1}{k_i})^2}\right],
\end{equation}
where the function $c(k) \defeqinl \tfrac{\Gamma(1+\nicefrac{2}{k})}{\Gamma(1+\nicefrac{1}{k})^2}$ is decreasing in $k$, with $c(\infty)=1$ and $c(1)=2$. Thus, for a non-``clumpy'' population---i.e., $p_\theta(k_i \geq 1) = 1$---the error is bounded by 2.
\end{prop}

The derivation of the estimator and error bound are given in~\apporweb{subsec:app_model}{Supplementary Material~A.1}. We note that while the estimator only depends on $\{s^{\msp{\tau}}, r^{\msp{\tau}}, s^{\msp{\lambda}}, r^{\msp{\lambda}}\}$, the error bound depends only on $\{s^{\msp{k}}, r^{\msp{k}}\}$, which is a consequence of the mean-concentration parameterization we have adopted.

\textbf{Incorporating covariates $\btheta(\boldsymbol{x}_i)$.} In the next section, we derive a variational inference scheme that enables gradient-based parameter estimation for WTYD. Beyond its scalability, a benefit of this approach is that it makes it easy to modify the WTYD model to incorporate covariates so that the parameters in~\cref{eq:wtyddeath,eq:wtyd_kprior,eq:wtyd_lamprior} become heterogeneous $\btheta_i \defeqinl \{s_i^{\msp{\tau}}, r_i^{\msp{\tau}}, s_i^{\msp{\lambda}}, r_i^{\msp{\lambda}}, s_i^{\msp{k}}, r_i^{\msp{k}}\}$, and defined as the output $\btheta_i \leftarrow \btheta(\boldsymbol{x}_i)$ of some differentiable function $\btheta(\cdot)$---e.g., a feedforward neural network---whose parameters we leave implicit. This is analogous to approaches for incorporating covariates into the classic BTYD models, described in~\cref{eq:covariates}, which has been difficult to apply to non-Poisson models when MCMC is used for inference.

%% file: sections/5_inference.tex
\section{Amortized Variational Inference for WTYD}
\label{sec:inference}

The marginal likelihood $p_{\btheta}(\bD_{1:n}) = \prod_i \int p_{\btheta}(\mathcal{D}_i, \bz_i)\, \textrm{d}\bz_i$ of the model in the previous section is intractable, and thus parameter estimation cannot proceed by directly maximizing it. This section develops an approach based on variational inference (VI) that is highly scalable and exploits modern machine learning frameworks for GPU-accelerated computation, automatic differentation, and neural network architectures. For background on the statistical foundations of VI and modern advances, see~\citet{blei2017variational} and \citet{zhang2018advances}, respectively.

\subsection{Variational EM with amortized Weibull factors}
The marginal likelihood (or ``\textit{evidence}'') $p_{\btheta}(\bD_{1:n}) = \prod_i \int p_{\btheta}(\mathcal{D}_i, \bz_i)\, \textrm{d}\bz_i$ is intractable under the model and thus cannot be maximized directly. Instead, we take a variational inference approach which seeks to maximize the following \textit{evidence lower bound (ELBO)}:
\begin{equation}
\mathcal{B}(\bphi, \btheta) \,\,\defeq\,\, \mathbb{E}_{\bz_{1:n} \sim q_{\bphi}}\left[\log \frac{p_{\btheta}(\bD_{1:n}, \bz_{1:n})}{q_{\bphi}(\bz_{1:n} \mid \bD_{1:n})}\right] \,\,\leq\,\, \log p_{\btheta}(\bD_{1:n}),
\label{eq:elbo}
\end{equation}
where $q_{\bphi}(\bz_{1:n} \mid \bD_{1:n})$ is the \textit{variational distribution}, parameterized by parameters $\bphi$. For any setting of model parameters $\btheta$, maximizing the ELBO with respect to $\bphi$ is equivalent to minimizing the KL divergence from the variational distribution to the exact posterior---i.e., $\mathrm{KL}(q_{\bphi}(\bz_{1:n} \mid \bD_{1:n}) \,\|\, p_{\btheta}(\bz_{1:n}\mid\bD_{1:n}))$. Thus, the fitted variational distribution  approximates the posterior.

Iteratively maximizing the ELBO with respect to $\bphi$ and $\btheta$ constitutes an extension of the original expectation-maximization (EM) algorithm~\citep{dempster1977maximum} known as \textit{variational EM}, which is guaranteed to converge to a local optimum of the ELBO~\citep{neal1998view,wainwright2008graphical}. This guarantee is retained when generalizing to take gradient steps~\citep{wu1983convergence}---i.e.,
\begin{align}
\label{eq:vestep_main}
\bphi &\leftarrow \textsc{Update}\big(\bphi,\, \nabla_{\bphi} \mathcal{B}(\bphi, \btheta)\big) \hspace{-3em}&&(\textrm{variational E-step}) \\
\label{eq:gmstep_main}
\btheta &\leftarrow \textsc{Update}\big(\btheta,\, \nabla_{\btheta} \mathcal{B}(\bphi, \btheta)\big) \hspace{-3em}&&(\textrm{generalized M-step})
\end{align}
where both steps are often performed simultaneously. While optima of the ELBO do not always match optima of the marginal likelihood, variational EM has been shown to yield consistent parameter estimates in a range of models~\subcitep[e.g.,]{titterington2006convergence}{hall2011asymptotic}, with extensive success reported for others; we find similar success in the next section.~\looseness=-1

\parhead{Variational family.} To facilitate optimization, we take a standard mean-field approach under which the variational distribution factorizes across all latent variables---i.e.,
\begin{equation*}
q_{\bphi}(\bz_{1:n}\mid\bD_{1:n}) = \prod_{i=1}^n q_{\bphi_i}(k_i\mid\mathcal{D}_i)\,\, q_{\bphi_i}(\lambda_i\mid\mathcal{D}_i)\,\, q_{\bphi_i}(\tau_i\mid\mathcal{D}_i).
\end{equation*}
We further define the family for all three customer-specific \textit{factors} to be Weibull, as follows:
\begin{align}
\label{eq:qk}
q_{\bphi_i}(k_i\mid\mathcal{D}_i) &= \textrm{Weibull}\big(\textsc{shape}\!=\!\tilde{s}_i^{\mathsmaller{\msp{k}}},\, \textsc{rate}\!=\!\tilde{r}_i^{\mathsmaller{\msp{k}}}\big), \\
\label{eq:qlam}
q_{\bphi_i}(\lambda_i\mid\mathcal{D}_i) &= \textrm{Weibull}\big(\textsc{shape}\!=\!\tilde{s}_i^{\mathsmaller{(\lambda)}},\, \textsc{rate}\!=\!\tilde{r}_i^{\mathsmaller{(\lambda)}}\big), \\
\label{eq:qtau}
q_{\bphi_i}(\tau_i\mid\mathcal{D}_i) &= \textrm{Weibull}\big(\textsc{shape}\!=\!\tilde{s}_i^{\mathsmaller{(\tau)}},\, \textsc{rate}\!=\!\tilde{r}_i^{\mathsmaller{(\tau)}},\, \textsc{location}\!=\!t_{i,y_i}\big),
\end{align}
where each customer is associated with variational parameters $\bphi_i \defeqinl \{\tilde{s}_i^{\mathsmaller{\msp{k}}}, \tilde{r}_i^{\mathsmaller{\msp{k}}}, \tilde{s}_i^{\mathsmaller{(\lambda)}}, \tilde{r}_i^{\mathsmaller{(\lambda)}}, \tilde{s}_i^{\mathsmaller{(\tau)}}, \tilde{r}_i^{\mathsmaller{(\tau)}}\}$ that define the shapes and rates of their three Weibull factors. The extra location parameter in~\cref{eq:qtau} assures that $\tau_i > t_{i,y_i}$---i.e., that customer $i$ was alive through their latest observed transaction---which has probability 1 under the exact posterior. The choice of Weibull for the factors will be key to low-variance gradient estimators, which we develop later.

\textbf{Amortization.} 
Fitting $\bphi_{1:n}$ for all $n$ customers is prohibitive for large datasets. However, we can ``\textit{amortize}'' the cost of inference~\citep{gershman2014amortized,zhang2018advances} by defining each customer-specific set of variational parameters to be the output of some learnable function $\bphi(\cdot)$ of the data---i.e., $\mathcal{D}_i \mathsmaller{\stackrel{\bphi(\mathcal{D}_i)}{\longmapsto}} \bphi_i$. This \textit{amortized variational inference} approach was first developed for variational autoencoders (VAEs)~\citep{kingma2014auto,rezende2014stochastic} but has since been applied for large-scale inference in more tailored statistical models, such as topic models~\subcitep{srivastava2017autoencoding}{sridhar2022heterogeneous}. An amortized variational distribution generally yields a worse approximation to the exact posterior---a tradeoff known as the ``amortization gap''~\subcitep{cremer2018inference}{margossian2024amortized}---but one that is tractable to compute for large datasets.~\looseness=-1

\textbf{Recognition network.} Parameterizing $\bphi(\cdot)$ as a neural network, it is common to refer to it as the \textit{``recognition (or inference) network''}. The variational parameters that are fit during inference are then the weights and biases of this network.\footnote{In a slight abuse of notation, we will use $\bphi$ to refer to both the neural network map and its parameters.} If the network were defined as a standard feedforward neural network, it would require a fixed-length input. However, in our case, the input data $\mathcal{D}_i$ includes the variable-length set of ITTs $\boldsymbol{\itt}_i$, since they do not have a sufficient statistic representation under the model. One option is to follow recent work on \textit{neural point processes}~\subcitep[e.g.,]{mei2017neural,shchur2021neural}{zuo2020transformer,zhou2022neural} in defining $\bphi(\cdot)$ to be an architecture tailored to sequences, such as a recurrent neural network (RNN) or a transformer. However, a simpler and less ``data hungry'' approach is to prespecify ``pseudo-sufficient'' statistics---i.e., a function $u(\cdot)$ mapping data to a fixed-length feature representation $\mathcal{D}_i \mathsmaller{\stackrel{u(\mathcal{D}_i)}{\longmapsto}} \bu_i$. A heuristic that we find works well is defining $u(\cdot)$ to be an expansion of the sufficient statistics for the Pareto/GGG model---i.e.,
\begin{equation}
    \label{eq:recognition}
\mathcal{D}_i \!\defeq\!
\begin{bmatrix*}[c]
    \itt_{i,1}\\
    \vdots \\
    \itt_{i,y_i}\\
    T_i
\end{bmatrix*} \!\in\! \mathbb{R}^{y_i+1} \hspace{1.5em}\mathlarger{\mathlarger{\stackrel{u(\mathcal{D}_i)}{\longmapsto}}}\hspace{1.5em} \boldsymbol{u}_i \!\defeq\!
\begin{bmatrix*}[c]
    y_i\\
    t_{i,y_i}\\
    \kappa_i\\
    T_i\\
    \zeta^{\msp{1}}_i\\
    \vdots\\
    \zeta^{\msp{P}}_i
\end{bmatrix*} \!\in\! \mathbb{R}^{P+4} \hspace{1.5em}\mathlarger{\mathlarger{\stackrel{\bphi(\bu_i)}{\longmapsto}}}\hspace{1.5em} \bphi_i \!\defeq\! \begin{bmatrix*}[l]
           \tilde{s}_i^{\msp{\tau}} \\
           \tilde{r}_i^{\msp{\tau}} \\
           \tilde{s}_i^{\msp{\lambda}} \\
           \tilde{r}_i^{\msp{\lambda}}\\
           \tilde{s}_i^{\msp{k}} \\
           \tilde{r}_i^{\msp{k}}
\end{bmatrix*} \!\in\! \mathbb{R}^{6},
\end{equation}
where $\zeta_i^{\msp{p}} \defeqinl \sum_{j} (\itt_{i,j})^{\nicefrac{p}{10}}$ is the sum of fractional-powers of ITTs. The vector $\bu_i$ exactly matches the sufficient statistics for Pareto/GGG when $P\!=\!0$. We find good performance with $P\!=\!25$ and with $\bphi(\cdot)$ defined as a multilayer perceptron (MLP) with at least two hidden layers. More details can be found in~\apporweb{sec:app_details}{Supplementary Material~B}. We will sometimes refer to the composition of both maps in~\cref{eq:recognition} as the ``recognition network'' $\bphi(\mathcal{D}_i)\equiv \bphi(u(\mathcal{D}_i))$.

\subsection{Stochastic optimization with the Weibull ``reparameterization trick''}

Evaluating the exact gradients in~\cref{eq:vestep_main,eq:gmstep_main} involves computation across the entire dataset $\mathcal{D}_{1:n}$ which can be prohibitive when $n$ is large. This subsection derives unbiased gradient estimators---i.e., $\widehat{\nabla}_{\bphi} \mathcal{B}(\bphi, \btheta)$ and $\widehat{\nabla}_{\btheta} \mathcal{B}(\bphi, \btheta)$---that can be computed instead on minibatches $\mathcal{M} \subset \boldsymbol{\mathcal{D}}_{1:n}$. In general, gradient ascent with unbiased gradient estimates is guaranteed to converge to a local mode of the objective provided that the step-sizes meet the conditions of~\citet{robbinsmonro1951}, a fact that has inspired the use of stochastic optimization~\subcitep{hoffman2013svi}{bottou2018optimization} with a broader range of optimizers (e.g., ADAM).

In developing the stochastic optimization procedure in this section, it will be useful to define the ``\textit{local learning signal}'' as $b_i(\bz_i;\, \bphi, \btheta) \defeqinl \log \tfrac{p_\btheta(\mathcal{D}_i, \bz_i)}{q_\bphi(\bz_i \mid \mathcal{D}_i)}$ and ``\textit{local ELBO}'' as
\begin{align}
\mathcal{B}(\bphi, \btheta) =  \sum_{i=1}^n \mathcal{B}_i(\bphi, \btheta) \,\,\textrm{ where }\,\,  \mathcal{B}_i(\bphi, \btheta) &\defeq  \mathbb{E}_{\bz_i \sim q_{\bphi}(\bz_i \mid \mathcal{D}_i)}[b_i(\bz_i; \bphi, \btheta)].
\end{align}
Given an unbiased estimator for the gradient of the local ELBO $\widehat{\nabla} \mathcal{B}_i(\bphi, \btheta)$, it is straightforward to obtain an unbiased gradient estimate of the full ELBO from a minibatch $\mathcal{M} \sim \textrm{Uniform}(1\dots n)$, since $\nabla \mathcal{B}(\bphi, \btheta) = \sum_{i=1}^n \nabla \mathcal{B}_i(\bphi, \btheta)$ and thus $\tfrac{n}{|\mathcal{M}|}\sum_{i \in \mathcal{M}} \widehat{\nabla} \mathcal{B}_i(\bphi, \btheta)$ is unbiased for $\nabla \mathcal{B}(\bphi, \btheta)$. The problem thus reduces to deriving unbiased estimators for the local gradients.~\looseness=-1

\textbf{Estimating $\nabla_{\btheta}\mathcal{B}_i(\bphi, \btheta)$.} This case is simple, as the gradient pushes into the expectation:
\begin{equation}
\label{eq:gmstep_main_further}
\nabla_{\btheta} \mathcal{B}_i(\bphi,\, \btheta) = \nabla_{\btheta}\mathbb{E}_{\bz_i \sim q_{\bphi}(\bz_i \mid \mathcal{D}_i)}[b_i(\bz_i; \bphi, \btheta)] = \mathbb{E}_{\bz_i \sim q_{\bphi}(\bz_i \mid \mathcal{D}_i)}[\nabla_{\btheta} b_i(\bz_i; \bphi, \btheta)],
\end{equation}
An unbiased estimator is then straightforward to obtain via Monte Carlo:
\begin{equation}
\widehat{\nabla}^{\textrm{MC}}_{\btheta} \mathcal{B}_i(\bphi, \btheta) \,\,\defeq\,\, \tfrac{1}{S}\sum_{s=1}^S \nabla_\btheta b_i(\bz_{i,s}; \bphi, \btheta) \,\,\textrm{ where }\,\, \bz_{i,s} \iidsim q_\bphi(\bz_i \mid \mathcal{D}_i).
\label{eq:gmstep_main_mc}
\end{equation}
We also note that for this particular model $\nabla_{\btheta} b_i(\bz_i; \bphi, \btheta) = \nabla_\btheta \log p_\btheta(\bz_i)$. 

\textbf{Estimating $\nabla_{\bphi}\mathcal{B}_i(\bphi, \btheta)$.} In this case, the gradient does not push in because $\bphi$ parameterizes the distribution governing the expectation---i.e.,
\begin{align}
\label{eq:estep_main_further}
\nabla_{\bphi} \mathcal{B}_i(\bphi,\, \btheta) \,\,=\,\, \nabla_{\bphi}\mathbb{E}_{\bz_i \sim q_{\bphi}(\bz_i \mid \mathcal{D}_i)}[b_i(\bz_i; \bphi, \btheta)] \,\,\neq\,\, \mathbb{E}_{\bz_i \sim q_{\bphi}(\bz_i \mid \mathcal{D}_i)}[\nabla_{\bphi} b_i(\bz_i; \bphi, \btheta)],
\end{align}
Instead, we will exploit the following fact. A Weibull can be simulated as a deterministic function of unit Exponential noise---i.e.,
\begin{equation}
\label{eq:reparam_weibull_main}
w \sim \textrm{Weibull}(\textsc{shape}{=}s, \textsc{rate}{=}r, \textsc{loc}{=}\ell)\,\,\,\Longleftrightarrow\,\,\,w = \varepsilon^{\nicefrac{1}{s}} r^{-1}+ \ell \,\textrm{ where }\,\varepsilon\sim\textrm{Exp}(1).
\end{equation}
Simulating $\bz_i \defeqinl [k_i,\lambda_i,\tau_i] \sim q_{\bphi}(\bz_i \mid \mathcal{D}_i)$ can thus proceed by sampling three Exponential variables $\beps_i \defeqinl [\varepsilon_i^{\msp{k}},\varepsilon_i^{\msp{\lambda}},\varepsilon_i^{\msp{\tau}}] \iidsiminl \textrm{Exp}(1)$ and computing the function $\bz_i = \bz_{\bphi}(\beps_i;\mathcal{D}_i)$ defined as
\begin{equation}
\label{eq:zreparam}
\bz_{\bphi}(\beps_i;\mathcal{D}_i) \,\,\defeq\,\, \Bigg[\frac{(\varepsilon_i^{\msp{k}})^{1/\tilde s_i^{\msp{k}}}}{\tilde r_i^{\msp{k}}},\, \frac{(\varepsilon_i^{\msp{\lambda}})^{1/\tilde s_i^{\msp{\lambda}}}}{\tilde r_i^{\msp{\lambda}}},\, \frac{(\varepsilon_i^{\msp{\tau}})^{1/\tilde s_i^{\msp{\tau}}}}{\tilde r_i^{\msp{\tau}}} + t_{i,y_i}\Bigg],
\end{equation}
where the variational parameters are output from the recognition network $\bphi(\mathcal{D}_i)$ in~\cref{eq:recognition}. This allows us to rewrite $\nabla_\bphi \mathbb{E}_{\bz_i \sim q_\bphi(\bz_i \mid \mathcal{D}_i)}[f(\bz_i)] = \mathbb{E}_{\beps_i}[\nabla_\bphi f(\bz_\bphi(\beps_i; \mathcal{D}_i))]$ for any $f(\bz_i)$ since $\bphi$ no longer parameterizes the expectation, allowing $\nabla_\bphi$ to push in. Applying this to~\cref{eq:estep_main_further}, we can then obtain the following unbiased gradient estimator via Monte Carlo:
\begin{align}
\label{eq:estep_main_mc}
\widehat{\nabla}_{\bphi}^{\textrm{MC}} \mathcal{B}_i(\bphi, \btheta) \,\,\defeq\,\,
\tfrac{1}{S}\sum_{s=1}^S \nabla_{\bphi} b_i\big(\bz_\bphi(\beps_{i,s}; \mathcal{D}_i);\,\bphi, \btheta\big) \,\,\textrm{ where }\,\, \beps_{i,s} \iidsim \textrm{Exp}(1).
\end{align}
This idea was first developed as the  ``pathwise gradient'' estimator~\citep{henderson2006handbooks} and later re-introduced as the ``reparameterization trick'' for inference in variational autoencoders (VAEs) with Gaussian latent variables~\subcitep{kingma2014auto,rezende2014stochastic}{titsias2014doubly}. Not all parametric families can be reparameterized explicitly, such as the Gamma distribution, for which approximate schemes for reparameterization gradients are required~\citep{figurnov2018implicit}. The Weibull is one of the few non-Gaussian distributions for which the explicit ``reparameterization trick'' is available, a fact which was exploited first by~\citet{zhang2018whai} to construct a non-negative VAE, and later by a handful of others~\subcitep[e.g.,]{squires2019variational,wang2024scalable}{chen2020switching,duan2021sawtooth,wang2021multimodal,zhang2021bayesian}.

\begin{algorithm}[t!]
\DontPrintSemicolon
\SetAlgoNoEnd
\SetInd{0.25em}{1.5em}
\caption{Amortized VI for WTYD with \optional[teal!100!black]{improvements and optional extensions in green}}
\label{alg:algorithm}
\KwIn{training data $\bD_{1:n}$; $\textsc{update}(\cdot)$; minibatch size $|\mathcal{M}|$; samples $S$; noise bounds $[a,b]$;\\\hspace{3em}initialized parameters $\btheta$, recognition network $\bphi(\cdot)$; \optional[teal!100!black]{(optional) synthetic data size $n_0$};\\\hspace{3em}\optional[teal!100!black]{(optional) validation data: $\bD_{n+1:n+V}$ (observation period), $\bD^\star_{n+1:n+V}$ (holdout)}}
\vspace{0.25em}\KwOut{fitted parameters $\btheta$, recognition network $\bphi(\cdot)$}

\vspace{0.5em}\optional[teal!100!black]{%
\tcc{(Optional) Inference compilation, \cref{subsec:details}}
$(\hat{\mathcal{D}}_i, \hat{\bz}_i)_{i=1}^{n_0} \iidsiminl p_{\btheta}(\mathcal{D}_i, \bz_i)$ \acom{sample synthetic dataset of size $n_0$}
\RepUntil(\acomf{fit network $\bphi(\cdot)$ in supervised manner}){maximum iteration}{
    $\bphi \leftarrow \textsc{Update}\big(\bphi,\, \textstyle\nabla_{\bphi} \log q_{\bphi}(\hat{\bz}_{1:n_0} \mid \boldsymbol{\hat{\mathcal{D}}}_{1:n_0})\big)$
}}

\vspace{0.5em}
\RepUntil{ELBO converges}{\label{line:trainloop_start}
    $\mathcal{M} \sim \textrm{Uniform}\big(\{1,\dots,n\}\big)$ \acom{sample minibatch of size $|\mathcal{M}|$}
    \For{$i \in \mathcal{M},\, s = 1,\dots,S$}{
        $\beps_{i,s} = [\varepsilon_{i,s}^{\msp{k}},\varepsilon_{i,s}^{\msp{\lambda}},\varepsilon_{i,s}^{\msp{\tau}}] \iidsiminl \textrm{TruncExp}_{[a,b]}(1)$ \acom{sample truncated Exponential noise}
        $\bz_{i,s} = [k_{i,s}, \lambda_{i,s}, \tau_{i,s}] \leftarrow \bz_{\bphi}(\beps_{i,s}; \mathcal{D}_i)$ \acom{Weibull reparameterization, \cref{eq:zreparam}}
    }
    \vspace{0.4em}$\widehat{\mathcal{B}}^{\textrm{RB}}(\bphi,\btheta) \leftarrow \tfrac{n}{|\mathcal{M}|}\sum\limits_{i\in\mathcal{M}} \widehat{\mathcal{B}}^{\textrm{RB}}_i(\bphi,\btheta)$ \acom{Rao--Blackwellized ELBO,~\cref{eq:elbo_rb}}
    $\bphi \leftarrow \textsc{Update}\big(\bphi,\ \nabla_{\bphi}\widehat{\mathcal{B}}^{\textrm{RB}}(\bphi, \btheta)\big)$ \acom{E-step (gradient via backpropagation)}
    $\btheta \leftarrow \textsc{Update}\big(\btheta,\ \nabla_{\btheta}\widehat{\mathcal{B}}^{\textrm{RB}}(\bphi, \btheta)\big)$ \acom{M-step (gradient via backpropagation)}\label{line:trainloop_end}

    \vspace{0.5em}
    \optional[teal!100!black]{    
    \tcc{(Optional) Early stopping, \cref{subsec:details}}
    Estimate $q_{\bphi}(\mathcal{D}^\star_{n+1:n+V} \mid \mathcal{D}_{n+1:n+V})$ \acom{posterior predictive density,~\cref{eq:postpred_mc}}
    \vspace{0.25em}\If{no improvement in recent iterations}{set $\btheta,\bphi(\cdot)$ to best checkpoint and \textbf{stop}}
    }
}
\vspace{0.55em}\Return $\btheta, \bphi(\cdot)$
\end{algorithm}

\parhead{Basic algorithm.} 
The core procedure is given in~\Crefrange{line:trainloop_start}{line:trainloop_end} of \cref{alg:algorithm}. Rather than explicitly implement the gradient estimators in~\cref{eq:gmstep_main_mc,eq:estep_main_mc}, we compute a Monte Carlo estimate of the ELBO $\widehat{\mathcal{B}}^{\textrm{MC}}(\bphi, \btheta)$ and obtain both gradient estimates via backpropagation (i.e., one pass of reverse-mode automatic-differentiation) which, since the reparameterization trick makes $\mathcal{B}(\bphi, \btheta)$ a differentiable function of both $\bphi$ and $\btheta$, is mathematically equivalent. As displayed, \cref{alg:algorithm} computes the Rao--Blackwellized refinement $\widehat{\mathcal{B}}^{\textrm{RB}}$ of this estimator, given in \cref{eq:elbo_rb} of \Cref{subsec:details}. 

\parhead{Software implementation.} The basic algorithm is easy to implement as both $p_\btheta$ and $q_\bphi$ involve only simple functions, all of which are supported for automatic differentiation in all major machine learning packages, such as \textsc{PyTorch}~\citep{ansel2024pytorch2}, which we use. We note that $p_\theta$ for the Pareto/GGG model involves the incomplete gamma function, which is not supported in most automatic differentiation frameworks, like \textsc{PyTorch}, and is thus why the change to a Weibull transaction process enables this form of scalable inference.

\subsection{Algorithmic improvements and practical details}
\label{subsec:details}
This section provides further refinements to the core variational algorithm that improve its practical performance, mainly by reducing the variance of gradient estimates and providing effective initialization to the recognition network. It also provides details on effective optimization schedules, with per-experiment settings further collected in \apporweb{sec:app_details}{Supplementary Material~B}. 

\parhead{Rao-Blackwellization.} The variance of the Monte Carlo ELBO estimate of \Cref{alg:algorithm}, and thus of the gradients backpropagated through it, can be reduced by exploiting semi-analytic structure that the ELBO admits. Specifically, we can decompose the learning signal into
\begin{align}
b_i(\bz_i;\, \bphi, \btheta) \,\,&=\,\, \underbrace{\log \tfrac{p_\btheta(k_i, \lambda_i)}{q_\bphi(\bz_i \mid \mathcal{D}_i)}}_{\defeq \,\, b_i^{\textrm{RB}}(\bz_i;\, \bphi, \btheta)} \,\,\,+\,\,\,  \underbrace{\log p(\mathcal{D}_i \mid \bz_i)p_\btheta(\tau_{i})}_{\defeq \,\, b_i^{\neg \textrm{RB}}(\bz_i;\, \bphi, \btheta)}, 
\end{align}
where we can show that the expectation of the first term is analytic. Modifying the algorithm so it computes the exact expectation of the first term is a form of Rao-Blackwellization, which decreases variance~\citep{ranganath2014black}. The Rao-Blackwellized estimator of the local ELBO is then~\looseness=-1
\begin{align}
\label{eq:elbo_rb}
\widehat{\mathcal{B}}^{\textrm{RB}}_i(\bphi, \btheta) &\defeq \mathbb{E}_{\bz_i \sim q_\bphi(\bz_i \mid \mathcal{D}_i)}[b_i^{\textrm{RB}}(\bz_i;\, \bphi, \btheta)] + \tfrac{1}{S}\sum_{s=1}^S b_i^{\neg \textrm{RB}}(\bz_{i,s};\, \bphi, \btheta),
\end{align}
where the first expectation, which can be formed analytically, equals
\begin{align}
\mathbb{E}_{\bz_i \sim q_\bphi(\bz_i \mid \mathcal{D}_i)}[b_i^{\textrm{RB}}(\bz_i;\, \bphi, \btheta)] &= \mathbb{H}\left[q_{\bphi}(\tau_i\mid\mathcal{D}_i)\right] - \mathrm{KL}\big(q_{\bphi}(k_i, \lambda_i \mid\mathcal{D}_i)\,\,\,\|\,\,\,p_{\btheta}(k_i, \lambda_i)\big).
\end{align}
where $\mathbb{H}$ is Shannon entropy, which is $\mathbb{H}[\textrm{Weibull}(\tilde s,\tilde r, \ell)] = \gamma (1- \nicefrac{1}{\tilde s}) - \log(\tilde{s}\tilde{r}) + 1$ for a Weibull. Note that it does not depend on location $\ell$. The KL term is the sum of two KL divergences, each from a Weibull variational factor to a Gamma prior, the general form for which follows from the KL between two Generalized Gamma distributions~\citep{bauckhage2014computing}---i.e., $$\mathrm{KL}(\textrm{Weibull}(\tilde s,\tilde r) \,\|\, \textrm{Gamma}(s,r)) = \log(\tilde{s}\,\Gamma(s)) + s\log (\nicefrac{\tilde r}{r}) + (\nicefrac{r}{\tilde r})\Gamma(1+\nicefrac{1}{\tilde s}) + \gamma(\nicefrac{s}{\tilde s}-1) -1.$$ Both expressions are differentiable in $\tilde s,\tilde r$, which correspond to the variational parameters, and contribute zero Monte Carlo variance to the gradient when computed exactly.

\parhead{Inference compilation.}
We initialize the recognition network by fitting it in a supervised manner to pairs $(\hat{\mathcal{D}}_i, \hat{\bz}_i) \,\mathsmaller{\iidsim}\, p_\btheta(\mathcal{D}_i, \bz_i)$ simulated from the generative model for a fixed initial $\btheta$. In other words, we initialize $\bphi$ by maximizing the conditional log-likelihood of simulated data:~\looseness=-1
\begin{align}
\max_{\bphi} \, \mathbb{E}_{(\mathcal{D}_i, \bz_i) \sim p_{\btheta}(\mathcal{D}_i, \bz_i)}\left[{\log\,} q_\bphi(\bz_i \mid \mathcal{D}_i)\right] 
\end{align}
The variational family $q_\bphi(\bz_i\mid\mathcal{D}_i)$ is thus initialized to a good approximation of the exact posterior $p_{\btheta}(\bz_i \mid \mathcal{D}_i)$ for the initial $\btheta$. It therefore produces sensible samples of $\bz_i$ when beginning the main training loop which then updates $\btheta$ and $\bphi$ jointly. This procedure is cheap, as it only requires sampling synthetic data and doing gradient ascent over a fixed dataset. In the Bayesian literature, this idea is known as \textit{``inference compilation''}~\citep{le2017inference}. It also corresponds to the sleep phase of the ``wake-sleep'' algorithm~\subcitep{hinton1995wake}{bornschein2015reweighted,le2020revisiting} and is an instance of ``forward amortized VI''~\citep{ambrogioni2019forward}. 

\parhead{Implicit gradient clipping via noise truncation.} When the variational shape parameter $\tilde s$ is small, the reparameterization map $\varepsilon \mapsto \varepsilon^{\nicefrac{1}{\tilde s}}/\tilde r$ of~\cref{eq:zreparam} can yield an extreme sample of $\bz_i$. Extreme values can cause numerical instability and large spikes in the training loss which hinder convergence. We resolve this by replacing the unit exponential noise with a \emph{truncated} exponential over $[a,b]$ with $a=\log(\nicefrac{1}{0.99})\approx 0.01$ and $b=\log(\nicefrac{1}{0.01})\approx 4.6$, discarding only the most extreme $1\%$ of each tail. This acts as an adaptive gradient-clipping scheme, which can be coupled with more standard gradient clipping. This trades gradient variance against a small amount of reparameterization bias. A similar scheme is reported by~\citet{zhang2018whai}.

\parhead{Early stopping.} A small fraction of customers can be held out during training as a validation set to assess overfitting. For $V$ of such customers, we split their ITT histories into observation and holdout periods so that the validation data consists of $\mathcal{D}_{n+1:n+V}$ and $\mathcal{D}^\star_{n+1:n+V}$. During training, we then estimate the posterior predictive density $q_{\bphi}(\mathcal{D}^\star_{n+1:n+V} \mid \mathcal{D}_{n+1:n+V})$ under the variational distribution (see~\Cref{subsec:var_approx}), tracking its running maximum. If a certain number of consecutive iterations fail to increase this maximum, we stop the training loop early and return the most recent checkpoint.

\parhead{Learning-rate schedule.} We use AdamW \citep{loshchilovhutter2019adamw} for the gradient $\textsc{Update}$ in \cref{alg:algorithm} with a learning rate following a cosine schedule annealed to near zero, with no warmup or restarts. Per-experiment initial rates, batch sizes, and network widths, among other details are given in \apporweb{sec:app_details}{Supplementary Material~B}.

\subsection{Variational approximations of the posterior quantities}
\label{subsec:var_approx}
The three posterior key quantities listed in \Cref{sec:background} are analytically intractable under the model $p_\btheta$. However, the variational posterior $q_{\bphi}$ has convenient analytic properties that enable principled approximations to all three. This section provides an overview of those approximations, with derivations and proofs relegated to \apporweb{subsec:app_inference}{Supplementary Material~A.2}.

\parhead{$P(\textsc{alive})$.}  Replacing the intractable posterior marginal $p_\btheta(\tau_i\mid\mathcal{D}_i)$ by the Weibull variational factor $q_\bphi(\tau_i\mid\mathcal{D}_i)$ and integrating over $(T_i,\infty)$ yields a closed-form approximation---i.e.,
\begin{equation}
\label{eq:palive_var}
q_\bphi(\textsc{alive}\mid\mathcal{D}_i) \,\,\defeq\,\, \int_{T_i}^\infty q_\bphi(\tau_i\mid\mathcal{D}_i)\,d\tau_i = \exp\Big(-\big[\tilde r_i^{\msp{\tau}}(T_i - t_{i,y_i})\big]^{\tilde s_i^{\msp{\tau}}}\Big).
\end{equation}
which is a simple function of variational parameters, recency, and total time.

\parhead{Posterior predictive distribution.}
Given a training window $(0,T_i]$ and a test window $(T_i,T_i^\star]$ with held-out future TTs $\mathcal{D}_i^\star$, the posterior predictive density is $p_\btheta(\mathcal{D}_i^\star\mid\mathcal{D}_i) = \mathbb{E}_{p_\btheta(\bz_i\mid\mathcal{D}_i)}[p(\mathcal{D}_i^\star\mid\bz_i)]$. Here and below, $p(\mathcal{D}_i^\star\mid\bz_i)$ suppresses an implicit conditioning on $\mathcal{D}_i$, which enters through the time already elapsed since the last observed transaction (see \apporweb{subsec:app_inference}{Supplementary Material~A.2}).  The plug-in variational approximation is then
\begin{align}
\label{eq:post_pred_approx}
\nonumber q_\bphi(\mathcal{D}_i^\star\mid\mathcal{D}_i) \,\,&\defeq\,\, \mathbb{E}_{q_\bphi(\bz_i\mid\mathcal{D}_i)}[p(\mathcal{D}_i^\star\mid\bz_i)]\\
&= q_\bphi\big(\tau_i > t^\star_{i,y_i^\star}\mid\mathcal{D}_i\big)\, \cdot \mathbb{E}_{q_\bphi(\bz_i\mid\mathcal{D}_i,\, \tau_i > t^\star_{i,y_i^\star})}\big[p(\mathcal{D}_i^\star\mid\bz_i)\big],
\end{align}
which factors cleanly into a survival probability and a density conditional on survival. The first factor takes the form of $q_\bphi(\textsc{alive}\mid\mathcal{D}_i)$ in~\cref{eq:palive_var} while the second is the individual-level likelihood in~\cref{eq:indivlikelihood} evaluated on $\mathcal{D}_i^\star$ and averaged over the variational posterior $q_\bphi(\bz_i\mid\mathcal{D}_i,\, \tau_i > t^\star_{i,y_i^\star})$. The following is thus a Monte Carlo estimator
\begin{equation}
    \label{eq:postpred_mc}
\widehat{q}_\bphi^{\,\textrm{MC}}(\mathcal{D}_i^\star\mid\mathcal{D}_i) \defeq q_\bphi(\tau_i > t^\star_{i,y_i^\star}\mid\mathcal{D}_i) \,\cdot \left[\frac{1}{S}\sum_{s=1}^S p(\mathcal{D}_i^\star\mid\bz_i^{(s)})\right],\quad \bz_i^{(s)} \iidsim q_\bphi(\bz_i\mid\mathcal{D}_i, \tau_i > t^\star_{i,y_i^\star}).
\end{equation}
which only relies upon drawing i.i.d.~samples of $\bz_i$ from the conditional variational posterior. Since all three latent variables are independent under $q_{\bphi}$, this amounts to drawing $k_i^{\mathsmaller{(s)}}$ and $\lambda_i^{\mathsmaller{(s)}}$ from their Weibull factors and $\tau_i^{\mathsmaller{(s)}}$ from its Weibull factor left-truncated at $t^\star_{i,y_i^\star}$.

\parhead{Forecasted future frequency.} The forecast frequency is the posterior, finite-horizon counterpart of the expected CLV derived in \Cref{sec:weityd_dev}. The plug-in variational approximation is
\begin{equation}
\label{eq:forecast_var}
\mathbb{E}_{\bphi}\left[Y_i(T_i, T_i^\star) \mid \mathcal{D}_i\right]
\,\,\defeq\,\, \mathbb{E}_{\bz_i \sim q_\bphi(\bz_i \mid \mathcal{D}_i)}\Big[\mathbb{E}\left[Y_i\big(T_i,\, T_i^\star \wedge \tau_i\big) \,\big|\, \bz_i\right]\Big].
\end{equation}
where the inner expectation is that of a Weibull Count random variable which can be Monte Carlo approximated or computed exactly via the recurrence of~\citet{mcshane2008count}. The posterior analogue to the Monte Carlo estimator for expected CLV in~\cref{eq:clv_mc} is thus
\begin{equation}
\label{eq:forecast_approx_mc}
\widehat{\mathbb{E}}^{\textrm{MC}}_{\bphi}\left[Y_i(T_i, T_i^\star) \mid \mathcal{D}_i\right]
\,\,\defeq\,\, \frac{1}{S}\sum_{s=1}^S \mathbb{E}\left[Y_i\big(T_i, T_i^\star \wedge \tau_i^{(s)}\big) \,\big|\, \bz_i^{(s)}\right], \quad \bz_i^{(s)}\iidsim q_\bphi(\bz_i\mid\mathcal{D}_i).
\end{equation}
We can also derive an analytic approximation, analogous to~\cref{eq:clv_rt}, which uses the same renewal theorem~\citep{feller1971} and the factorization of $q_\bphi$. For any $T_i^\star > T_i$ it is
\begin{equation}
\label{eq:forecast_approx_rt}
\widehat{\mathbb{E}}^{\textrm{RT}}_{\bphi}\left[Y_i(T_i, T_i^\star) \mid \mathcal{D}_i\right]
\;\defeq\; \mathbb{E}_{\bphi}[\lambda_i]\,\cdot\,q_\bphi(\textsc{alive}\mid\mathcal{D}_i)\,\cdot\,\mathbb{E}_{\bphi}\left[T_i^\star \wedge \tau_i - T_i \,\big|\, \tau_i > T_i\right], 
\end{equation}

where all three factors are closed-form functions of the variational parameters, as detailed in \apporweb{subsec:app_inference}{Supplementary Material~A.2}. As with CLV, we can further bound the bias of this approximation using a similar argument as used in \Cref{prop:clv_bias}.

\begin{prop}[Renewal-theoretic bias bound]
\label{prop:rt_bias}
For any $T_i^\star > T_i$, the renewal-theoretic estimator in \cref{eq:forecast_approx_rt} of the variational forecast frequency in \cref{eq:forecast_var} satisfies 
\begin{equation}
\label{eq:rt_bias_bound}
\big|\widehat{\mathbb{E}}^{\textrm{RT}}_{\bphi}\left[Y_i(T_i,T_i^\star)\mid\mathcal{D}_i\right] - \mathbb{E}_{\bphi}\left[Y_i(T_i,T_i^\star)\mid\mathcal{D}_i\right]\big|
\;\leq\; q_\bphi\left(\textsc{alive}\mid\mathcal{D}_i\right)\cdot \mathbb{E}_{\bphi}\left[\frac{\Gamma(1+\nicefrac{2}{k_i})}{\Gamma(1+\nicefrac{1}{k_i})^2}\right].
\end{equation}
where the function $c(k) \defeqinl \tfrac{\Gamma(1+\nicefrac{2}{k})}{\Gamma(1+\nicefrac{1}{k})^2}$ is decreasing in $k$, with $c(\infty)=1$ and $c(1)=2$. Thus, for a non-``clumpy'' customer---i.e., $q_\bphi(k_i \geq 1 \mid \mathcal{D}_i) = 1$---the error is bounded above by 2 events.
\end{prop}

%% file: sections/6_simulations.tex

\section{Simulation Studies}
\label{sec:performance_eval}

We first evaluate the proposed model and variational algorithm with a series of synthetic-data studies. \Cref{subsec:submodels} assesses the variational approximation in a simplified setting where the exact posterior is known.  \Cref{subsec:param_recovery,subsec:latent_recovery} then assess recovery of ground-truth $\btheta$ and $\bz_{1:n}$, respectively, comparing the variational WTYD approach to Pareto/GGG with MCMC. \Cref{subsec:scaling} gives a systematic comparison across different synthetic datasets of the held-out predictive performance of WTYD and Pareto/GGG as a function of wall-clock time.

\subsection{Recovery of conjugate posteriors in simplified sub-model}
\label{subsec:submodels}

\begin{figure}[t!]
    \centering
    \includegraphics[width=\linewidth]{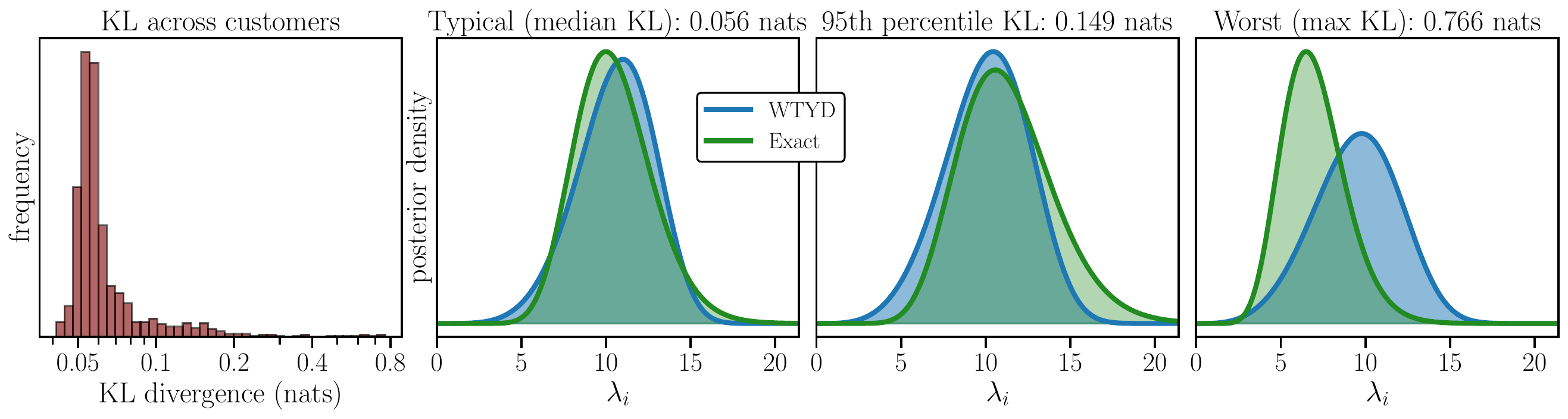}
    \caption{The Weibull variational posterior closely matches the exact Gamma posterior in the closed-form sub-model ($k_i=1$, observed $\tau_i$). The per-customer KL from the Weibull $q$ to the exact Gamma posterior is small across the population (median $0.056$ nats). The variational (WTYD) versus exact posteriors over $\lambda_i$ for a typical customer (median KL), the $95$th-percentile customer, and the worst case show that the two are nearly indistinguishable except in the tail, and only for the most extreme customers.}
    \label{fig:submodels}
\end{figure}

As an initial check to see whether an amortized VI approach can yield a good approximation to the posterior, we first generated synthetic data according to the Pareto/NBD model, with  $\lambda_i\sim\mathrm{Gamma}(s^{\msp{\lambda}},r^{\msp{\lambda}})$, $\tau_i\sim\text{Pareto}(s^{\msp{\tau}},r^{\msp{\tau}})$ and $k_i=1$. Each customer's transaction times were then sampled on the interval $(0, T_i \wedge \tau_i]$ according to a Poisson process, with the total time fixed at $T_i = 1$ for all customers. Conditional on $\tau_i$, the exact posterior over $\lambda_i$ in this case is known:~\looseness=-1
$$
p_\btheta(\lambda_i \mid \mathcal{D}_i,\tau_i, k_i=1) = \mathrm{Gamma}(\lambda_i;\, \textsc{shape}=s^{\msp{\lambda}}+y_i,\,\textsc{rate}=r^{\msp{\lambda}}+T_i \wedge \tau_i) 
$$
Conditioning on the true $\tau_i$ and $k_i=1$, we then fit the Weibull variational posterior $q_\bphi(\lambda_i \mid \mathcal{D}_i)$ using the procedure outlined in~\Cref{sec:inference}, with a recognition network $\bphi(\cdot)$ whose input is each customer's variable-length set of ITTs $\mathcal{D}_i = \{\boldsymbol{\Delta t}_i, T_i\}$. 

To evaluate the approximation, we use the KL divergence $\textrm{KL}(q_\bphi(\lambda_i \mid \mathcal{D}_i) \,\,\|\,\, p_\btheta(\lambda_i \mid \mathcal{D}_i,\tau_i, k_i=1))$ between the learned Weibull variational distribution and the true Gamma posterior, which can be computed in closed form. The results are given in~\Cref{fig:submodels} where the per-customer KL clusters around a median of $0.056$ nats. It is only for the very worst case ($0.766$ nats) where the variational approximation appears clearly wrong, with approximations at the $95^{\textrm{th}}$ percentile of KL (0.149 nats) still appearing reasonable visually.

\begin{figure}[t!]
    \centering
    \includegraphics[width=\linewidth]{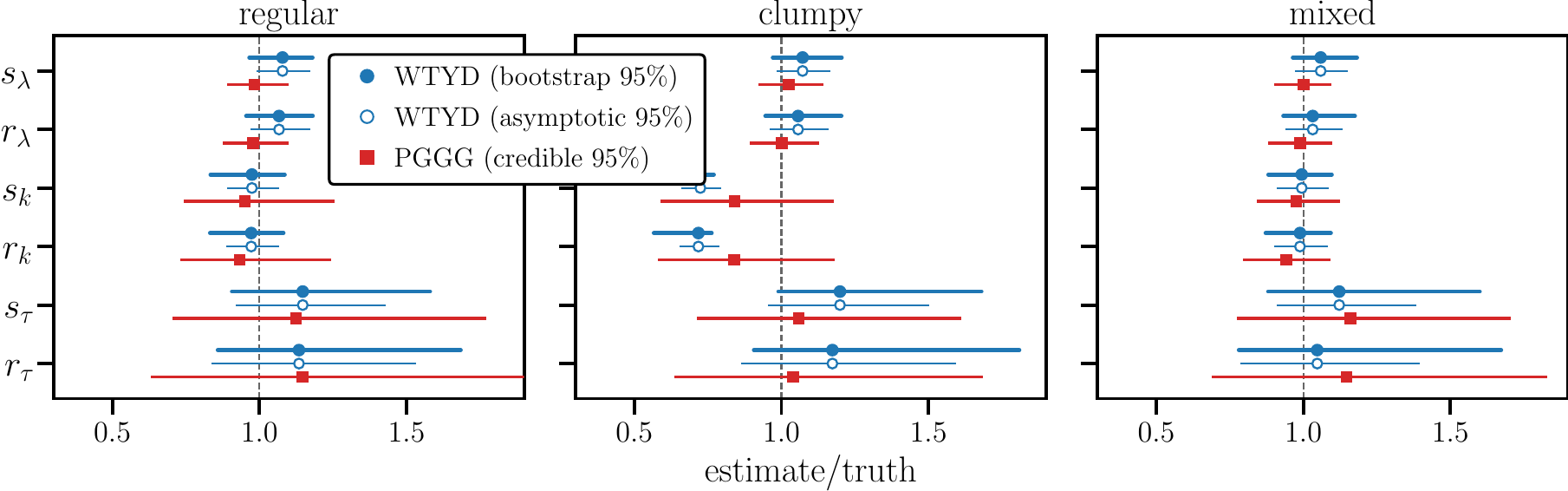}
    \caption{WTYD recovers known population hyperparameters and matches Pareto/GGG in the mixed and regular regimes. Each of the hyperparameters is plotted as $\mathrm{estimate}/\mathrm{truth}$, so the vertical line marks perfect recovery. For WTYD, we show both the customer-level bootstrap interval and the cheaper asymptotic interval; the Pareto/GGG reference is its posterior credible interval. Both models cover the generating values, except in the clumpy regime, where WTYD underestimates the regularity concentration.} 
    \label{fig:param_recovery_hetT} 
\end{figure}

\subsection{Population hyperparameter recovery}
\label{subsec:param_recovery}

We next ask whether variational WTYD recovers known population-level hyperparameters $\btheta = \{s^{\msp{\tau}}, r^{\msp{\tau}}, s^{\msp{\lambda}}, r^{\msp{\lambda}}, s^{\msp{k}}, r^{\msp{k}}\}$ as accurately as Pareto/GGG using MCMC as implemented by \citet{platzer2016customer}. We generated panels of $n=1{,}000$ customers, drawing $T_i \sim \textrm{Uniform}(1, 10)$, $\lambda_i \sim \textrm{Gamma}(2,\, \nicefrac{1}{5})$ and $\tau_i \sim \textrm{Pareto}(2,\, 5)$, and considering three regimes for regularity:
$$
\overset{\textrm{(``regular'')} \,\,\, \textrm{95\% ETI}=(1.62,\, 3.57)}{k_i \sim \textrm{Gamma}(25,\, 10)}, \hspace{1em}
\overset{\textrm{(``clumpy'')} \,\,\, \textrm{95\% ETI}=(0.63,\, 1.02)}{k_i \sim \textrm{Gamma}(65,\, 80)}, \hspace{1em}
\overset{\textrm{(``mixed'')} \,\,\, \textrm{95\% ETI}=(0.70,\, 3.26)}{k_i \sim \textrm{Gamma}(7,\, 4)}
$$
where each is annotated with its $95\%$ equal-tailed interval (ETI). The ``regular'' prior's density is concentrated above $1$, the ``clumpy'' prior below $1$, and the ``mixed'' prior evenly spread. 

Because the two models assume different renewal families, we generated each regime twice, with Weibull versus Gamma renewals, holding the true generated $\bz_{1:n}$ and $\btheta$ fixed, and fitting each model to the variant matching its own likelihood. We trained WTYD for a fixed budget of $80{,}000$ iterations with no early stopping, and fit Pareto/GGG using a single MCMC chain run for $3{,}100$ iterations, with the first $100$ discarded as burn-in and every $10$th iteration retained thereafter ($300$ draws); all remaining settings are given in \apporweb{tab:app_experiment_settings}{Supplementary Material~B}.

The results are shown in~\Cref{fig:param_recovery_hetT} where both models do comparably well in almost all settings. The only exception is recovering the regularity hyperparameters $\{s^{\mathsmaller{(k)}}, r^{\mathsmaller{(k)}}\}$ in the ``clumpy'' regime, which both models do poorly on, with WTYD's confidence intervals further failing to cover the true value, unlike the credible intervals of Pareto/GGG. This is consistent with variational approximations typically underestimating the posterior variance \citep{blei2017variational}.

We used two different methods to obtain confidence intervals for WTYD's estimates, both computed from the single observed dataset rather than across simulation replicates. The first is a customer-level bootstrap: we resampled the $n$ customers with replacement and refit WTYD from scratch, repeating this $100$ times. The second is an asymptotic approximation which treats $\widehat{\btheta}$ as an M-estimator maximizing the ELBO and estimates its covariance by the inverse Hessian at the fitted optimum. While the bootstrap requires full refits, the asymptotic interval requires only one Hessian evaluation; however, this comes at the cost of approximate intervals which often underestimate those of the bootstrap. See~\apporweb{sec:app_details}{Supplementary Material~B} for more details.

\begin{figure}[t!]
    \centering
    \includegraphics[width=\linewidth]{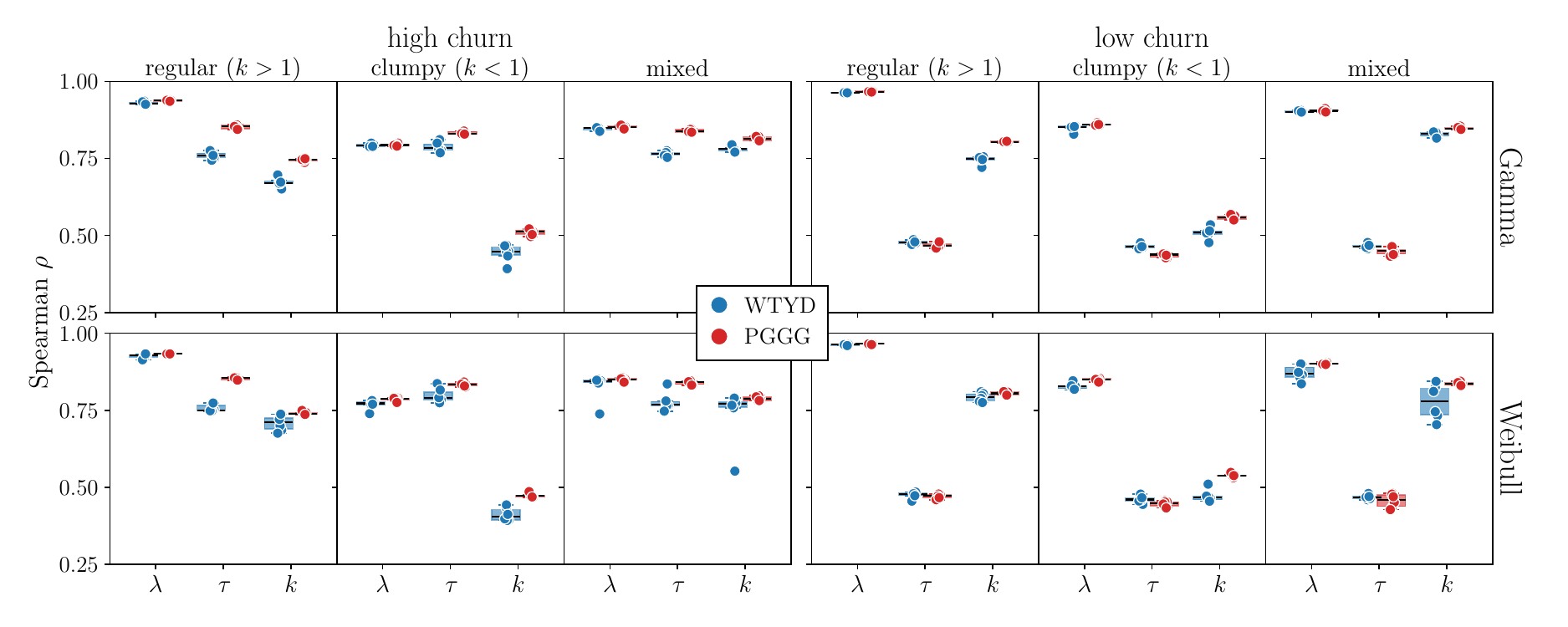}
    \caption{WTYD recovers customer-level latents as well as Pareto/GGG. Rank correlation between true and inferred $\lambda_i$, $k_i$, and $\tau_i$ across the synthetic grid. Lifetime recovery degrades when most customers remain alive at $T_i$ (no dropout signal); regularity recovery is hardest in the clumpy, high-churn corner.}
    \label{fig:rank_corr_n10000_simulation_results}
\end{figure}

\subsection{Customer-specific latent variable recovery}
\label{subsec:latent_recovery}

We next assess whether variational WTYD recovers the customer-specific latents $\bz_i = \{\tau_i, \lambda_i, k_i\}$ as accurately as Pareto/GGG across a range of data settings. We generate twelve settings defined by the cross-product of renewal family, regularity regime and churn level:\vspace{0.5em}
$$\{\textrm{Gamma},\textrm{Weibull}\} \times \{\textrm{``regular''}, \textrm{``clumpy''}, \textrm{``mixed''}\} \times \{\textrm{``low''}, \textrm{``high''}\}$$
where the ``high'' churn level draws $\tau_i \sim \textrm{Pareto}(2,\, 2.5)$ so that roughly $\nicefrac{1}{2}$ of customers drop out within the observation window, and ``low'' draws $\tau_i \sim \textrm{Pareto}(2,\, 8.5)$ so that roughly $\nicefrac{1}{5}$ drop out. Unlike the previous study, the regularity regimes here draw $k_i$ from bounded ranges: ``regular'' draws $k_i \sim \textrm{Uniform}(1.5,\, 10)$, ``clumpy'' draws $k_i \sim \textrm{Uniform}(0.5,\, 0.95)$, and ``mixed'' draws $k_i$ from an equal-parts mixture of the ``regular'' and ``clumpy'' distributions and a point mass at $k_i = 1$. These ranges are stated in units of the Gamma concentration where for the Weibull datasets we use the correspondence map of \Cref{prop:correspondance_prop}. All settings share $\lambda_i \sim \textrm{Gamma}(4,\, \nicefrac{3}{20})$ and $T_i = 1$. For each setting, we generate six datasets of $n = 10{,}000$, and fit both models to each, so that each model is evaluated under both matched and mismatched renewal families.~\looseness=-1

Since the scales of the latent variables $\{\lambda_i, k_i, \tau_i\}$ may change depending on the renewal family, we assess recovery using Spearman rank correlation between the ground-truth latent variable and the inferred posterior or variational mean---e.g., for $\tau_i$:
$$
\rho_\tau \,\defeq\, \textrm{corr}\big(\textrm{rank}(\tau_{1:n}),\, \textrm{rank}(\widehat{\tau}_{1:n})\big) \,\,\,\textrm{where}\,\,\,
\widehat{\tau}_i \,\defeq\, \mathbb{E}\left[\tau_i \mid \mathcal{D}_i\right]
$$
where this is an analytic expectation $\smash{\mathbb{E}_{q_{\bphi}(\tau_i \mid \mathcal{D}_i)}[\tau_i \mid \mathcal{D}_i]}$ for WTYD and an empirical average of MCMC samples \smash{$\tfrac{1}{S}\sum_{s=1}^S \tau_i^{\mathsmaller{(s)}}$} for Pareto/GGG. While the exact scales may differ, we expect both models to rank order customers correctly for all three variables.

\Cref{fig:rank_corr_n10000_simulation_results} visualizes the results where both models perform comparably. Pooled across settings and random seeds, the median correlation for WTYD versus Pareto/GGG is $0.867$ versus $0.881$ for $\lambda_i$, $0.725$ versus $0.765$ for $k_i$, and $0.617$ versus $0.655$ for $\tau_i$. We do not expect WTYD to recover latents \textit{better} than Pareto/GGG, as both models are equally expressive (\Cref{prop:correspondance_prop}), and amortized VI introduces approximation error. However, the gap between models is much smaller than the variation across settings, suggesting that the approximation error introduced by VI is minor. For instance, recovery of $\tau_i$ is largely governed by churn level for both models with $\rho_\tau$ falling from $0.76$--$0.86$ under ``high'' churn to $0.45$--$0.48$ under ``low'' churn, where $80\%$ of customers outlive the window and thus reveal little about their lifetimes. Similarly, recovery of $k_i$ is worst in the ``clumpy'' and ``high''-churn setting---i.e., $0.42$ for WTYD, $0.48$ for Pareto/GGG---where short, irregular histories contain few ITTs.

\begin{figure}[t!]
    \centering
    \includegraphics[width=\linewidth]{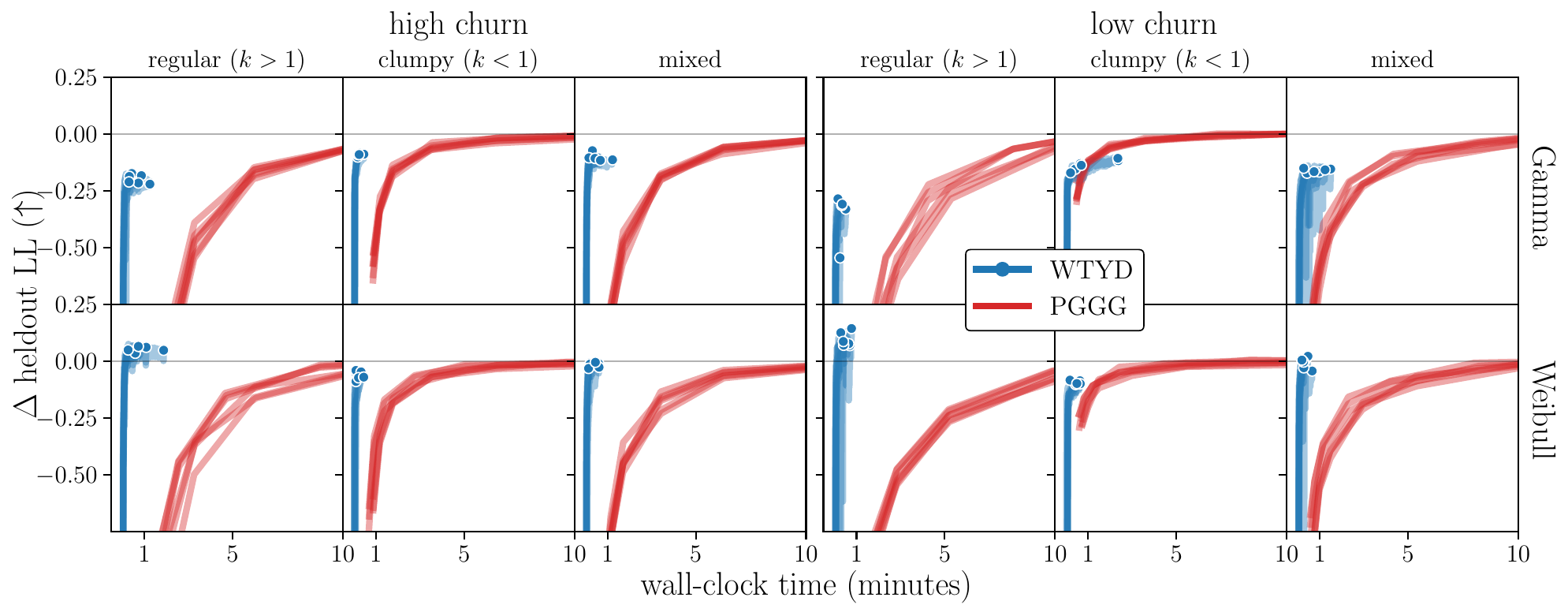}
    \caption{WTYD obtains comparable predictive performance at a fraction of Pareto/GGG's computational cost. Each curve shows the held-out posterior predictive density for WTYD (\textcolor{WTYDColor}{blue}) and Pareto/GGG (\textcolor{PGGGColor}{red}) over wall-clock training time on one of the six synthetic datasets we generated for each setting. We subtract from both curves the final value of Pareto/GGG, so that all red curves end at zero, and blue curves above-vs-below zero denote WTYD obtaining better-vs-worse predictive performance than Pareto/GGG at convergence.}
    \label{fig:higher_n_simulation_results}
\end{figure} 

\subsection{Predictive performance by wall-clock time}
\label{subsec:scaling}

The results of the preceding simulation studies suggest that variational WTYD and Pareto/GGG with MCMC perform comparably, with WTYD incurring minor approximation error. Here we assess the tradeoff that VI offers by measuring the two models' predictive performance as a function of wall-clock time. We re-use the synthetic datasets created in the previous study, this time further creating future holdout periods on which to assess forecasting performance. We again fit both models to each training dataset, this time tracking an estimate of the posterior predictive density of the held-out data. For WTYD, this is simply $q_\bphi(\mathcal{D}^\star_{i} \mid \mathcal{D}_{i})$ as estimated by~\cref{eq:postpred_mc} at each iteration, while for Pareto/GGG we use the accumulated set of MCMC samples at each iteration to estimate $p_\btheta(\mathcal{D}^\star_{i} \mid \mathcal{D}_{i})$ by averaging the analogous per-draw held-out likelihood; see \apporweb{subsec:app_inference}{Supplementary Material~A.2} for more details. When reporting wall-clock time, we do not include the time spent on computing these quantities.

\Cref{fig:higher_n_simulation_results} visualizes the results. We see that at this size $n{=}10{,}000$, WTYD typically converges before the MCMC chain for Pareto/GGG has burned in. At convergence, the performance of the two models is broadly comparable with small differences being consistent with renewal family mismatch as WTYD typically converges to better performance on the Weibull datasets and Pareto/GGG better on the Gamma datasets.

%% file: sections/7_case_study.tex
\section{Case Studies}
\label{sec:case_studies}

This section presents two case studies applying WTYD to predict and explore structure in two large-scale real-world datasets of 1) online retail transactions at an anonymous subscription-based firm whose CRM data we obtained from the company Ocurate and 2) political donations during the run-up to the 2020 U.S.~Presidential Election as recorded by the U.S. Federal Election Commission (FEC). These case studies both highlight the need for scalable non-Poisson models that can account for heterogeneity in regularity.

\input{sections/7a_retail}
\input{sections/7b_fec_donors}

%% file: sections/7a_retail.tex
\subsection{Case Study 1: 5 Million Subscription-Based Online Retail Customers}
\label{subsec:case_study_retail}

\begin{figure}[t!]
    \centering
    \includegraphics[width=\linewidth]{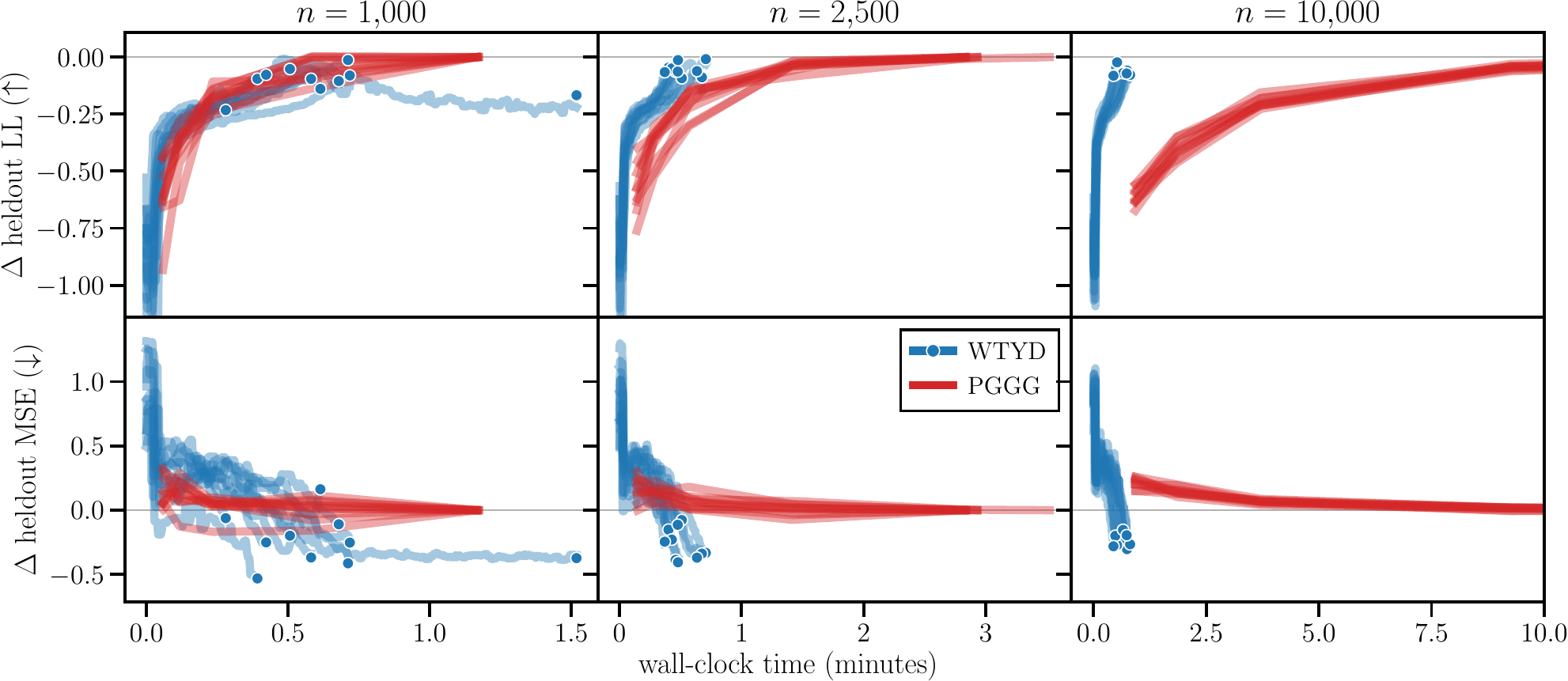}
    \caption{WTYD matches Pareto/GGG's forecast performance at a fraction of its wall-clock cost on real data. Following the conventions of \Cref{fig:higher_n_simulation_results}, each curve shows a model's performance relative to Pareto/GGG's final value over wall-clock training time, here on random retail subsets of size $n \in\{1000,2500,10000\}$ (columns). The top row reports the held-out posterior predictive density and the bottom row the mean-squared error of the forecasted future frequency, for which \textcolor{WTYDColor}{blue} curves below zero denote WTYD forecasting more accurately than Pareto/GGG at convergence.}
    \label{fig:forecast_panel_retail}
\end{figure}

Much of the literature on BTYD models focuses on Poisson models for the non-contractual setting. However, such models often fail to appropriately capture regularity in the contractual setting, which has been the primary motivation for the few non-Poisson models that have thus far been introduced~\citep[e.g.,][]{herniter1971probablistic,chatfield1973consumer,schmittlein1983prediction,platzer2016ticking,reutterer2021leveraging}. This case study centers around a subscription-based online retail firm, where many customers are contractually prompted to transact monthly or bimonthly, but can skip or delay transactions, and can otherwise transact at will, yielding ITT histories that exhibit regularity but are not trivially predictable.~\looseness=-1

\parhead{Data and preprocessing.}  We obtained anonymized CRM data from the marketing analytics company Ocurate\footnote{Ocurate has since been acquired by Fenix Commerce: \url{https://fenixcommerce.com/}.}. The dataset records 20.8 million transactions from 5.1 million unique customers over the course of five years from January 1, 2019 to February 14, 2024. As is typical, there is a high churn rate with a heavy tail of loyal customers, with 40\% of customers never making any repeat purchases, 25\% making five or more, and 2\% making 20 or more.

We processed the data into the standard ``customer's journey'' representation (\cref{fig:customer_journey}), shifting and rescaling each customer's history so that all customers are ``born'' at $t\!=\!0$ and are observed for different total times $T_i$. For the purpose of predictive evaluation, we also created training-holdout splits with two-thirds of each customer's true observation window designated for training $\mathcal{D}_i$ and the remaining held-out $\mathcal{D}^\star_i$.~\looseness=-1

\begin{figure}[t!]
    \centering
    \includegraphics[width=\linewidth]{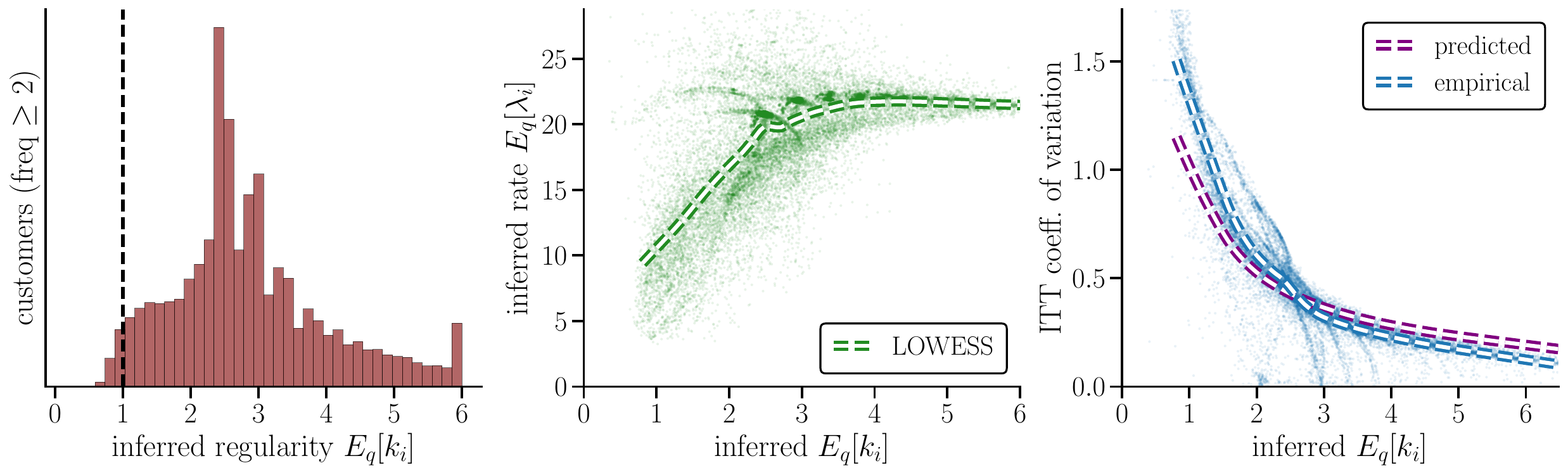}
    \caption{Inferred regularity structure on the full customer base. 
    \textbf{Left:}~The distribution of inferred $k_i$, where the dashed line marks $k_i \!=\! 1$, shows most customers are ``regular''.
    \textbf{Middle:}~The joint distribution of inferred $k_i$ and $\lambda_i$ with a LOWESS summary shows that more regular customers are also more frequent.
    \textbf{Right:}~The joint distribution of $k_i$ against the empirical coefficient of variation of ITTs $\widehat{\textrm{CV}}_i$ then shows high correspondence with the model's predicted relationship, suggesting that inferred relationship between regularity and frequency is not driven by model or approximate inference artifacts.}
    \label{fig:retail_exploratory}
\end{figure}

\parhead{Predictive evaluation.} We first repeat the same kind of forecasting experiments reported in~\Cref{subsec:scaling}, this time on real-world data, comparing the performance of variational WTYD to Pareto/GGG as a function of wall-clock time. Since MCMC with Pareto/GGG does not scale computationally to the size of the full dataset, we selected random subsets of $n \in\{1000,2500,10000\}$ customers, splitting each into a train-holdout split as described above. Both models were fit to each dataset using the same settings as in~\Cref{subsec:scaling}, computing the posterior predictive density of held-out data over training time. In addition, we also calculated the mean-squared error (MSE) of the true versus forecasted future frequency $\tfrac{1}{n}\sum_{i=1}^{n} (y_i^\star - \mathbb{E}[y_i^\star \mid \mathcal{D}_i])^2$ using the variational approximation of the forecasted future frequency $\mathbb{E}[y_i^\star \mid \mathcal{D}_i]$ given in \Cref{eq:forecast_approx_mc} for WTYD, and using Monte Carlo averages for Pareto/GGG---i.e., by simulating a Gamma renewal process at each posterior sample of $\boldsymbol{z}_i$. An estimator caveat affecting roughly $1\%$ of customers per subset in Pareto/GGG's held-out means is noted in \apporweb{sec:app_details}{Supplementary Material~B}.~\looseness=-1

\Cref{fig:forecast_panel_retail} visualizes the posterior predictive and mean-squared error results for all three dataset sizes as a function of training time. For the smallest size of $n=\textrm{1,000}$, both models converge comparably quickly to very similar values of both held-out posterior predictive density and mean-squared error. However, as the size increases to $n=$10,000, WTYD converges  before Pareto/GGG has even burned in, just like in~\Cref{subsec:scaling}.  Moreover, at convergence, both models exhibit comparable forecasting performance as measured by posterior predictive density, but with WTYD further exhibiting substantially lower mean-squared error in its approximate frequency forecasts. We note that while Pareto/GGG's performance curve is already dominated by WTYD at this size, $n=$10,000 still represents only 0.2\% of all 5 million customers. While we estimate fitting Pareto/GGG to the whole dataset would take around 3-4 days, WTYD is fit in under 8 minutes.~\looseness=-1

\parhead{Exploratory analysis.} Fitting WTYD to the full dataset, we then explored the inferred latent structure, as given by the variational means of $\bz_i \defeqinl \{k_i, \lambda_i, \tau_i\}$. The first figure in \Cref{fig:retail_exploratory} shows the distribution of inferred regularity $k_i$, with a large peak around $k_i \!=\! 2.5$ and substantial mass extending even beyond $k_i > 5$. Only a very small fraction of 2.5\% customers are ``clumpy'' or ``random'' with $k_i \!\leq\! 1$, as would be expected in a contractual setting. 

\begin{figure}[t!]
    \centering
    \includegraphics[width=\linewidth]{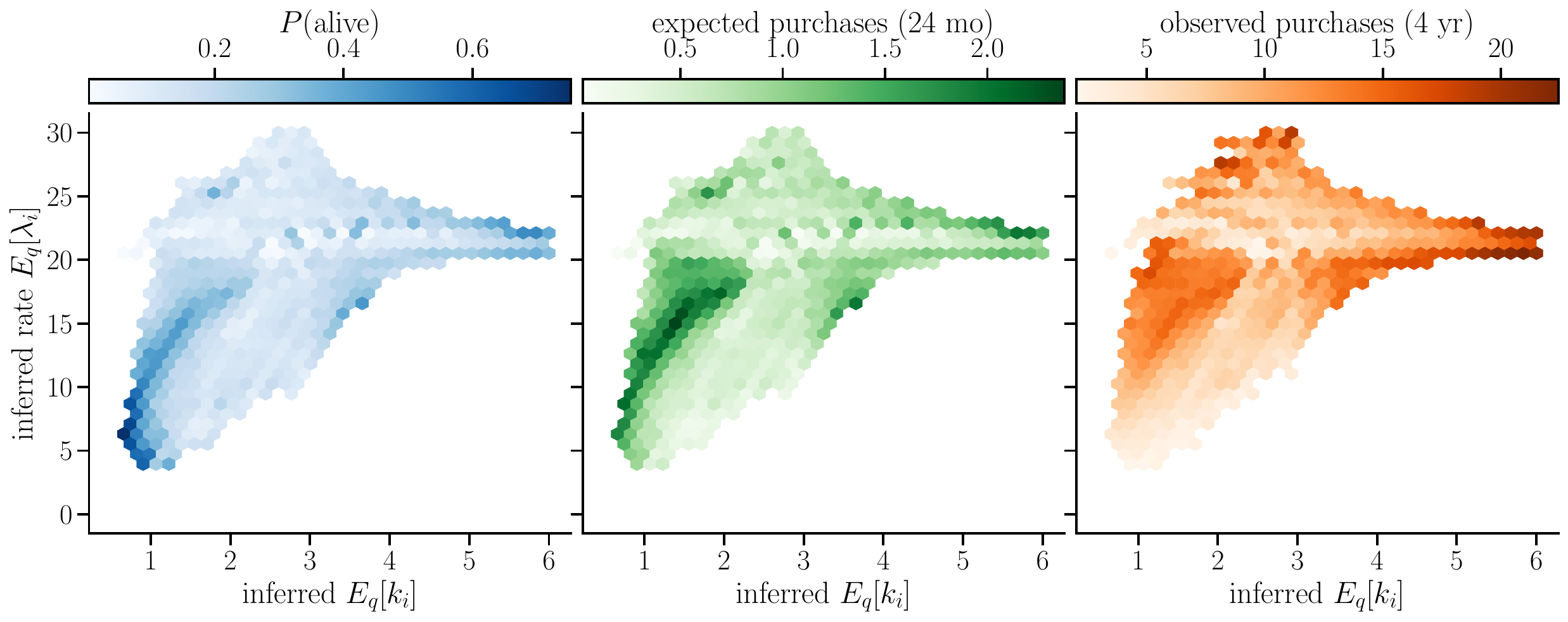}
    \caption{Inferred churn and expected future purchasing on the full customer base.
    \textbf{Left:}~The average of $q_{\bphi}(\textsc{alive} \mid \mathcal{D}_i)$ over the $(k_i, \lambda_i)$ plane shows that only a small segment of the base is inferred to be still alive, concentrated among ``low-low'' and ``high-high'' customers.
    \textbf{Middle:}~The expected number of purchases over the next 24 months shows that expected future purchasing concentrates among ``medium-medium'' customers.
    \textbf{Right:}~The observed frequencies of customers observed for at least four years also concentrate around ``medium-medium'' as well as ``high-high'' customers.}
    \label{fig:retail_churn}
\end{figure}

Although nothing in the model assumes this, we also find that more regular customers purchase more frequently.  The middle plot of \Cref{fig:retail_exploratory} shows the joint distribution of the inferred $k_i$ and $\lambda_i$ where increasing $k_i$ is highly predictive of larger $\lambda_i$ for $k_i \!<\! 4$ with the average transaction rate among ``random'' ($k_i=1$) customers being almost half that of more regular ones with $k_i \!=\! 3$. This relationship tapers off after $k_i \!>\! 4$ with transaction rates saturating at around $\lambda_i \!=\!20$. One might still wonder whether this inferred relationship is an artifact of approximate inference, and that larger $k_i$ is partly reflecting larger frequencies rather than simply the regularity of transaction timings. To examine this, we visualize in the rightmost plot of \Cref{fig:retail_exploratory} the joint distribution of $k_i$ with the empirical coefficient of variation of the customer's ITTs---i.e., $\widehat{\textrm{CV}}_i \defeqinl \nicefrac{s_i}{m_i}$, where $m_i$ and $s_i$ are the sample mean and standard deviation of customer $i$'s observed ITTs $\itt_{i,j}$, defined for customers with $y_i \!\geq\! 2$. The empirical relationship is just as predicted under the model, with larger $k_i$ strongly associated with smaller $\widehat{\textrm{CV}}_i$. We also overlay the theoretical relationship under the model, as given by $\textrm{CV}_i \defeqinl \sqrt{\nicefrac{\Gamma(1+\nicefrac{2}{k_i})}{\Gamma(1+\nicefrac{1}{k_i})^2} - 1}$, which aligns with a LOWESS summary of the empirical relationship.~\looseness=-1


We also explore the inferred structure of churn and expected future purchasing. The leftmost panel of \Cref{fig:retail_churn} visualizes average $q_{\bphi}(\textsc{alive} \mid \mathcal{D}_i)$ among subsets of customers with similar values of $\lambda_i$ and $k_i$. We see that only a small segment of the customer base has $q_{\bphi}(\textsc{alive} \mid \mathcal{D}_i) > 0.5$. This is concentrated in two regions: low frequency and low regularity (``low engagement''), where $\tau_i$ is most difficult to infer, and high frequency and high regularity (``high-high''), which comprises the firm's heaviest or most loyal customers.

We then compare this in the middle panel to the expected number of purchases over the next 24 months, computed using the estimator \Cref{eq:forecast_approx_rt}. Here the bulk of expected future purchases are concentrated among customers with medium frequency and medium regularity (``medium-medium''), who are less likely to be alive than low-low customers but transact faster when they are. We compare these forecasts in the rightmost panel to the observed empirical frequencies among the roughly 300,000 customers that were acquired early enough to be observed for at least four years. We see that the bulk of observed frequencies is also concentrated among the medium-medium customers, as well as the high-high, as expected.~\looseness=-1 

%% file: sections/7b_fec_donors.tex
\subsection{Case Study 2: 4 Million Donors in the 2020 US General Elections}
\label{subsec:fec}

This case study focuses on individual political contributions 2019--2020 during the run-up to the US General elections. In this setting, there is a mix of contractual and non-contractual contributions. Moreover, funding campaigns, deadlines, and other structural features modulate the timing of contributions, inducing a range of ``clumpy'' and ``regular'' timing patterns, as we will show. Political donation dynamics have previously been shown to respond to filing deadlines, debates, and other salient events~\subcitep{magleby2018donates,bouton2022small}{traag2016complex}, and non-political charitable giving has been studied as a canonical application of BTYD models~\subcitep{netzer2008hidden}{fader2010customer} including Pareto/GGG~\citep{platzer2016ticking,valendin2022customer}.

\begin{figure}[t!]
    \centering
    \includegraphics[width=\linewidth]{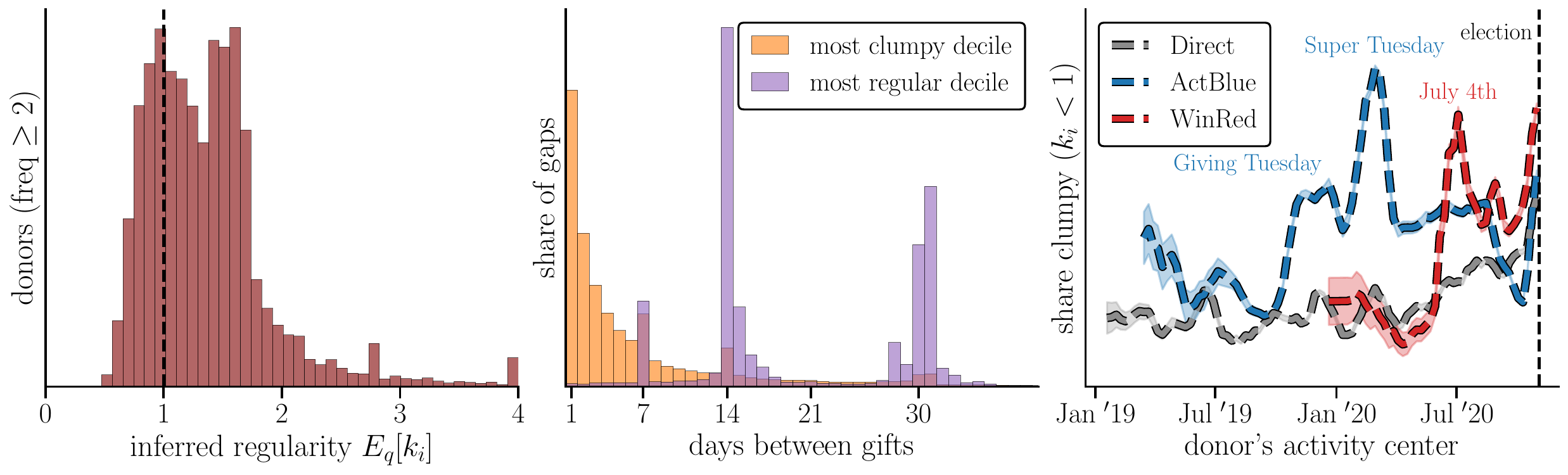}
    \caption{Inferred regularity structure on the full donor base.
    \textbf{Left:}~A bimodal distribution of inferred $k_i$.
    \textbf{Middle:}~The distribution of ITTs among the most and least ``regular'' decile of donors.
    \textbf{Right:}~Share of each week's donors that are ``clumpy'', smoothed by a 3-week moving average, with 95\% confidence intervals.}
    \label{fig:fec_exploratory}
\end{figure}

\parhead{Data and preprocessing.} We obtained data from the public bulk files of the U.S. Federal Election Commission (FEC)~\citep{fecbulk}, which record every itemized contribution by an individual to a federal committee and include the conduit records through which the partisan companies ActBlue and WinRed route small-dollar gifts. We filtered to individual donors with valid transaction codes, positive amounts, and gifts inside the cycle window from January 1, 2019 to election day, November 3, 2020, and further collapse multiple gifts on the same day into ``donation-days''. Since the FEC assigns no persistent donor identifier, we identify donors by hashed name and five-digit ZIP code. This results in a dataset of $24.5$ million donation-days from $4.1$ million unique donors over a window of $673$ days. The distribution of donor activity closely mirrors the retail panel in the previous case study with $40\%$ of donors giving on only a single day, $30\%$ giving on five or more days, and only $6\%$ on twenty or more~\looseness=-1.

\parhead{Exploratory analysis.} Fitting WTYD to the full dataset, we again explored the inferred latent structure. \Cref{fig:fec_exploratory} shows the distribution of inferred regularity $k_i$ where unlike in the retail panel where regularity dominated, the donor base here is bimodal, with one mode near $k_i=1$ and another near $k_i=1.75$. There is again a heavy right tail of highly ``regular'' donors and, unlike in the retail setting, a sizeable number of ``clumpy'' donors, accounting for nearly $30\%$ of the donor base. As in the previous case study, the inferred $k_i$ closely tracks the empirical coefficient of variation $\widehat{\textrm{CV}}_i$, with rank correlation $\rho \!=\! -0.79$, suggesting that the split reflects real differences in donors' timing patterns rather than artifacts of approximate inference. In the middle plot of \Cref{fig:fec_exploratory} we also show the distribution of ITTs among the most regular decile of donors where we see that $82\%$ of gaps fall at exactly $7$, $14$, $30$, or $31$ days, as would be expected under automated recurring contributions. 

The rightmost plot of \Cref{fig:fec_exploratory} then shows when in the cycle the most ``clumpy'' donors were giving. We assign each donor to the week containing the median date of their gifts, and then plot the share of each week's donors that are ``clumpy'', doing so separately for ActBlue, WinRed, and direct donors. Among ActBlue donors, this share spikes to $43\%$ around Giving Tuesday and to $69\%$ around Super Tuesday, the latter of which coincided with the largest single day of the 2020 Democratic presidential primary. Among WinRed donors, the share of ``clumpy'' instead spikes to around $60\%$ in the week spanning the end of the second fundraising quarter on June 30, a reporting deadline that campaigns target with a burst of appeals, and July 4th, a natural focal point for Republican fundraising. All three groups then also exhibit a spike in ``clumpiness'' just before Election Day, during which campaigns often concentrate appeals.

\begin{figure}[t!]
    \centering
    \includegraphics[width=\linewidth]{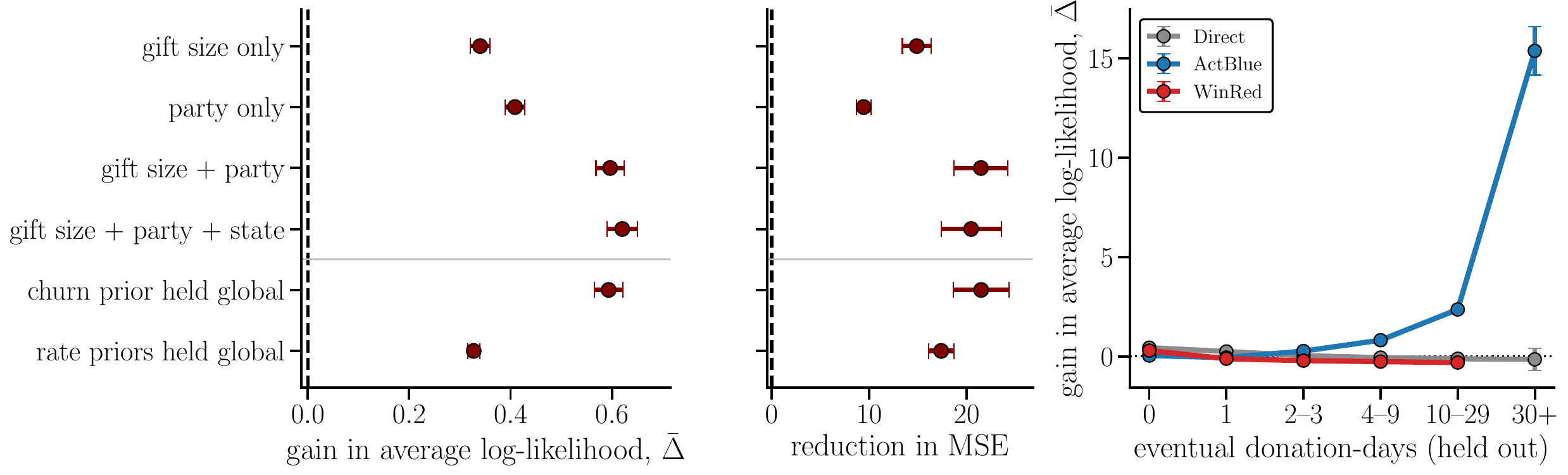}
    \caption{Cold-start prediction of held-out donors, where positive values favor the covariate prior.
    \textbf{Left:}~The gain in held-out log-likelihood of each covariate specification over the global prior.
    \textbf{Middle:}~The analogous reduction in squared error of the held-out frequency forecast.
    \textbf{Right:}~The gain log-likelihood gain by each donor's eventual number of donation-days, separately by conduit.}
    \label{fig:fec_coldstart}
\end{figure}

\parhead{Incorporating covariates for ``cold-start'' prediction.} We can associate with each donor a set of static covariates $\mathbf{x}_i$---i.e., party, US state, and first gift amount---where party is inferred from whether the donor gave through ActBlue ($\textrm{party}_i=\textsc{Dem}$), WinRed ($\textrm{party}_i=\textsc{Repub}$), or directly ($\textrm{party}_i=\textsc{None}$)\footnote{We note that WinRed launched in June 2019, partway through the observation window, and only records gifts above \$200. Thus, customers with $\textrm{party}_i = \textsc{Repub}$ are only those whose first gift is over \$200 and after June 2019.}.  Such covariates may be useful in improving ``cold start'' prediction for newly observed donors. As noted in \Cref{sec:weityd_dev}, WTYD's gradient-based inference algorithm makes incorporating covariates into the prior hyperparameters easy. We define the map to be log-linear---i.e., $\btheta(\mathbf{x}_i) \defeqinl \exp(W\mathbf{x}_i)$---where the exponential applies elementwise and $\mathbf{x}_i$ includes a constant. The model is then fit exactly as before, with gradients flowing through $\btheta(\mathbf{x}_i)$ to the coefficients $W$.~\looseness=-1

To see whether covariates improve ``cold start'' prediction, we constructed train-test splits at the donor level, holding out \textit{all} of the ITTs for 200,000 test donors and then fitting both the original and various covariate-conditional WTYD models to the remaining donor pool. We then evaluated the prior-predictive (i.e., marginal likelihood) density of the held-out donors' ITTs under the fitted models, along with the MSE of their true versus expected frequencies, as approximated by \cref{eq:clv_rt}.~\looseness=-1

\Cref{fig:fec_coldstart} visualizes the improvement in ``cold start'' prediction of various covariate-conditional models over the original. While conditioning on the US state of the donor yielded no improvement, both party and first gift size improve predictions independently. Moreover, we fit two further variants that localize covariates to specific parts of the model. In one case, the churn hyperparameters \smash{$\{s^{(\tau)}, r^{(\tau)}\}$} are kept global and not covariate-conditioned while in the other the frequency hyperparameters \smash{$\{s^{(\lambda)}, r^{(\lambda)}\}$} are kept global. While fixing the churn prior to be global has no effect, fixing the frequency prior substantially reduces the gain, suggesting that a donor's first gift and party is highly informative about how frequently they will donate when active, but carries little information about how long they will remain active.

The gains in performance are unevenly distributed across the donor pool, with the best-performing covariate-conditional model only improving in held-out log-likelihood for about two-thirds of test donors who are concentrated among more frequent ActBlue donors, as shown in the rightmost panel of \Cref{fig:fec_coldstart}. One explanation for why the impact of covariates is so uneven is that ActBlue donors are more heterogeneous, spanning the full range of giving styles, from large single-time gifts to sustained recurring schedules. For instance, $44\%$ of held-out ActBlue donors give on four or more donation-days compared to only $9.5\%$ of WinRed donors. First gift size is thus useful in predicting the subtype of ActBlue donor, with smaller first gifts suggesting a sustained recurring donor, while such information does not improve ``cold start'' prediction for the more homogeneous WinRed donor pool.~\looseness=-1

%% file: sections/8_conclusion.tex
\section{Discussion}
\label{sec:conclusion}

We found that heterogeneous regularity was indeed important to account for in two real-world settings of subscription-based retail and political donations, consistent with work over the last several decades that details departures from Poisson timing in transaction data~\citep{chatfield1973consumer,gupta1991stochastic,zhang2015predicting,platzer2016customer}. There are few existing non-Poisson BTYD models that account for such heterogeneity, and these models do not scale computationally to the size of modern datasets. Our approach breaks ground on this problem by pairing a BTYD model based on Weibull renewals with amortized variational inference. The proposed approach fits a dataset of 5 million customers in 8 minutes compared to an estimated 3-4 days that would have been required by the existing state-of-the-art, and does so without any appreciable loss in predictive performance or model interpretability.

Alongside a recent preprint by~\citet{naf2026clvae}, this paper is among the first to adapt amortized VI to BTYDs. \citeauthor{naf2026clvae}~introduce the Customer Lifetime Variational Autoencoder (CLVAE) which adapts amortized VI to fit Pareto/NBD. Adapting amortized VI beyond Pareto/NBD to non-Poisson models, the challenge that our paper addresses, is hampered by the form of the likelihood in more complex models, and specifically the inability to take gradients through it via automatic differentiation. In general, for any renewal family $f_i$, a BTYD likelihood takes the following form:
\begin{equation}
\label{eq:renewal_likelihood}
p(\mathcal{D}_i \mid \bz_i) \,\,=\,\, S_i\big((T_i \wedge \tau_i) - t_{i,y_i}\big) \prod_{j=1}^{y_i} f_i(\itt_{i,j}),
\end{equation}
where $S_i(\Delta t) \defeqinl \int_{\Delta t}^\infty f_i(u)\,\textrm{d}u$ is the survival function ---i.e., the probability of an ITT greater than $\Delta t$. For the Gamma renewals, $S_i(\Delta t) = \nicefrac{\Gamma(k_i,\, k_i\lambda_i \Delta t)}{\Gamma(k_i)}$, which involves the upper incomplete gamma function $\Gamma(s, x) \defeqinl \int_{x}^\infty u^{s-1} e^{-u}\, \textrm{d}u$, which is not supported in most automatic differentiation frameworks, like \textsc{PyTorch}. Replacing Gamma renewals with Weibull as we do then instead yields $S_i(\Delta t) = \exp(-(r_i \Delta t)^{k_i})$ and with it the likelihood in \cref{eq:indivlikelihood}, which involves only elementary functions beyond the gamma function, which all automatic differentiation frameworks support. 

In changing to a Weibull renewal model, we took care to adopt a mean--concentration parameterization which admits latent variables $\lambda_i$ and $k_i$ that retain their qualitative interpretation under Pareto/GGG, and moreover have a formal correspondence, as detailed in~\Cref{prop:correspondance_prop}. The two models are thus only very subtly different. In~\Cref{subsec:scaling}, each model attained slightly better held-out performance when the synthetic data were generated under its own renewal family, but with the gap between the two always being small. In~\Cref{subsec:case_study_retail}, WTYD seemed to fit the online retail data slightly better than Pareto/GGG. However, this was again with a small gap and moreover with there being no a priori reason to expect one model to fit better than the other. The main benefit is computational, with WTYD requiring orders of magnitude less time to fit, and with the additional benefit of the ease at which covariates can be incorporated, as illustrated in~\Cref{subsec:fec}.~\looseness=-1


A major tradeoff of the proposed approach is its lack of a sufficient statistic representation, which thus requires the recognition network to consume each customer's variable-length sequence of ITTs $\boldsymbol{\Delta t}_i$. We adopt a simple approach to this, mapping each sequence to a fixed-length vector of prescribed ``pseudo-sufficient'' statistics, as in \cref{eq:recognition}. A more sophisticated approach that future follow-up work could consider would define the recognition network to be a recurrent network or transformer, among other architectures tailored to sequences, drawing upon the emerging literature on neural point processes~\subcitep{mei2017neural,shchur2021neural}{zuo2020transformer,zhou2022neural}.~\looseness=-1

Future work might also seek to enrich the lifetime process. WTYD retains the simple Pareto lifetime process of its predecessors, but nothing in the inference procedure depends on this choice, as the lifetime density, like the renewal survival function in \cref{eq:renewal_likelihood}, need only be automatically differentiable. One avenue of future work that amortized VI thus enables is to parameterize the survival function directly as a deep neural network, drawing upon recent work on neural survival analysis~\subcitep[e.g.,]{wiegrebe2024deep}{katzman2018deepsurv}. One could also generalize beyond simple alive-versus-dead states via Markov-modulated renewal processes~\citep{pyke1961markov} such that customers transition through multiple transacting modes, drawing upon BTYD models inspired by hidden Markov models and other discrete-time processes~\citep[e.g.,][]{allenby1999dynamic,netzer2008hidden,ascarza2018some}. Amortized VI enables exploration of many such extensions as the only requirement it imposes is that the complete-data likelihood be automatically differentiable; any such model satisfying this can then be conceivably fit to large-scale datasets with covariates, as demonstrated here.

%% file: sections/appendix.tex
\appendix

\counterwithin*{table}{section}
\counterwithin*{figure}{section}
\renewcommand{\thetable}{\thesection.\arabic{table}}
\renewcommand{\thefigure}{\thesection.\arabic{figure}}
\renewcommand{\theHtable}{appendix.\thesection.\arabic{table}}
\renewcommand{\theHfigure}{appendix.\thesection.\arabic{figure}}

\section{Proofs and derivations}
\label{sec:app_proofs}

\subsection{Model properties and prior key quantities}
\label{subsec:app_model}

\parhead{Proof of~\Cref{prop:correspondance_prop} (correspondence of regularity index).}
In the mean--concentration parameterization the reverse KL from the Weibull (mean $\nicefrac{1}{\lambda_W}$, concentration $k_W$) to the Gamma (mean $\nicefrac{1}{\lambda_G}$, concentration $k_G$) is
\begin{align}
\label{eq:kl_gamma_wei_mean_concen}
\textrm{KL} = -\gamma - 1 + \log k_W - k_G\log(k_G\lambda_G) + \log\Gamma(k_G) + \tfrac{k_G\lambda_G}{\lambda_W} + k_G\log\!\big(\lambda_W\Gamma(1+\nicefrac{1}{k_W})\big) + \tfrac{k_G\gamma}{k_W},
\end{align}
where we used that the Weibull and Gamma rates in this parameterization are $\lambda_W\Gamma(1+\nicefrac{1}{k_W})$ and $k_G\lambda_G$. Setting $\partial/\partial\lambda_W=0$ gives $-k_G\lambda_G/\lambda_W^2 + k_G/\lambda_W = 0$, so $\lambda_W^\star=\lambda_G$; the second derivative $k_G/\lambda_G^2>0$ there confirms a minimum. Setting $\partial/\partial k_W=0$ gives
\begin{equation*}
k_G = k_W^\star\big(\gamma + \psi(1+\nicefrac{1}{k_W^\star})\big)^{-1},
\end{equation*}
which is~\cref{eq:kmap}. Since $\psi$ is increasing and $1+\nicefrac{1}{k_W}$ is decreasing in $k_W$, the denominator $\gamma+\psi(1+\nicefrac{1}{k_W})$ is decreasing, so the map $k_W^\star\mapsto k_G$ is strictly increasing (as is its inverse $k_G\mapsto k_W^\star$); at $k_W^\star=1$, $\psi(2)=1-\gamma$ gives $k_G=1$. Hence $k_W^\star$ and $k_G$ always lie on the same side of $1$. Uniqueness follows because the stationary point is unique---$\lambda_W^\star=\lambda_G$, with a unique $k_W^\star$ for each $k_G$ by strict monotonicity---while the KL in~\cref{eq:kl_gamma_wei_mean_concen} diverges as $k_W$ or $\lambda_W$ tends to $0$ or $\infty$, so this stationary point is the global minimum.

\parhead{Derivation of~\cref{eq:indivlikelihood} (individual-level likelihood).}
While alive, the ITTs are i.i.d.\ $\textrm{Weibull}(k_i,r_i)$ with $r_i=\lambda_i\Gamma(1+\nicefrac{1}{k_i})$, density $f(w)=k_i r_i^{k_i} w^{k_i-1} e^{-r_i^{k_i} w^{k_i}}$ and survival $S(w)=e^{-r_i^{k_i} w^{k_i}}$. The history $\mathcal{D}_i$ consists of the $y_i$ observed ITTs followed by a censored gap from the last transaction $t_{i,y_i}$ to the end of observation $T_i\wedge\tau_i$, during which no transaction occurs, so
\begin{equation*}
p(\mathcal{D}_i\mid\bz_i) = \Big[\textstyle\prod_{j=1}^{y_i} f(\itt_{i,j})\Big]\, S\big(T_i\wedge\tau_i - t_{i,y_i}\big).
\end{equation*}
Substituting $f$ and $S$ and collecting the exponentials yields~\cref{eq:indivlikelihood}; the indicator $\mathbb{1}(y_i>0)$ drops the transaction-density factor for never-purchasing customers, leaving the survival term alone.

\parhead{Derivation of~\cref{eq:clv_rt} (expected CLV estimator).}
The mean ITT is $\mu_i=\mathbb{E}[\itt_{i,j}]=\Gamma(1+\nicefrac{1}{k_i})/r_i=\nicefrac{1}{\lambda_i}$, so the long-run rate of the renewal process is exactly $\lambda_i$. By the elementary renewal theorem $\mathbb{E}[Y_i(t)\mid\bz_i]\to t/\mu_i=\lambda_i t$, hence $\mathbb{E}[Y_i(\tau_i)\mid\bz_i]\approx\lambda_i\tau_i$ for large $\tau_i$. Taking the prior expectation and using prior independence of $\lambda_i$ and $\tau_i$,
\begin{equation*}
\mathbb{E}_\btheta[Y_i(\infty)] \approx \mathbb{E}_\btheta[\lambda_i\tau_i] = \mathbb{E}_\btheta[\lambda_i]\,\mathbb{E}_\btheta[\tau_i] = \tfrac{s^{\msp{\lambda}}}{r^{\msp{\lambda}}}\cdot\tfrac{r^{\msp{\tau}}}{s^{\msp{\tau}}-1},
\end{equation*}
the product of the $\textrm{Gamma}(s^{\msp{\lambda}},r^{\msp{\lambda}})$ and $\textrm{Pareto}(s^{\msp{\tau}},r^{\msp{\tau}})$ prior means, which is~\cref{eq:clv_rt}. (Throughout, $\textrm{Pareto}(\textsc{shape}{=}s,\,\textsc{scale}{=}r)$ denotes the Pareto type-II, or Lomax, distribution with density $\tfrac{s}{r}(1+\nicefrac{\tau}{r})^{-(s+1)}$ on $\tau>0$, whose mean $\tfrac{r}{s-1}$ requires $s>1$.)

\parhead{Proof of~\Cref{prop:clv_bias} (error bound of CLV estimator).}
Lorden's upper bound on the renewal function~\citep{lorden1970excess}, combined with the elementary lower bound $\mathbb{E}[Y_i(t)\mid\bz_i]\ge t/\mu_i-1$, gives $\big|\mathbb{E}[Y_i(t)\mid\bz_i]-t/\mu_i\big|\le \mathbb{E}[\itt_{i,j}^2\mid\bz_i]/\mu_i^2$ for all $t$. For the Weibull ITT, $\mathbb{E}[\itt_{i,j}^2 \mid \bz_i]/\mu_i^2=\Gamma(1+\nicefrac{2}{k_i})/\Gamma(1+\nicefrac{1}{k_i})^2=:c(k_i)$, which is scale-free (independent of $r_i$). Evaluating at $t=\tau_i$, where $t/\mu_i=\lambda_i\tau_i$, and taking the prior expectation gives
\begin{equation*}
\big|\mathbb{E}_\btheta[Y_i(\infty)] - \widehat{\mathbb{E}}^{\textrm{RT}}_\btheta[Y_i(\infty)]\big| \le \mathbb{E}_{p_\btheta(k_i)}\!\big[c(k_i)\big],
\end{equation*}
which is~\cref{eq:clv_bias_bound}; the bound depends only on $\{s^{\msp{k}},r^{\msp{k}}\}$ because $c$ is scale-free. Because $c(k)$ grows super-exponentially as $k\to0$, the expectations in~\cref{eq:clv_bias_bound,eq:rt_bias_bound} are finite only when the distribution of $k_i$ is supported away from zero; if $k_i\ge k_0$ almost surely, monotonicity of $c$ gives the finite bound $c(k_0)$.

\subsection{Variational inference and posterior key quantities}
\label{subsec:app_inference}

\parhead{Derivation of~\cref{eq:palive_var} ($q_\bphi(\textsc{alive}\mid\mathcal{D}_i)$).}
$q_\bphi(\textsc{alive}\mid\mathcal{D}_i)$ is the variational probability that $\tau_i>T_i$. The $\tau$-factor is $\textrm{Weibull}(\tilde s_i^{\msp{\tau}},\tilde r_i^{\msp{\tau}},\textsc{loc}=t_{i,y_i})$, whose survival for $\tau>t_{i,y_i}$ is $\exp\!\big(-[\tilde r_i^{\msp{\tau}}(\tau-t_{i,y_i})]^{\tilde s_i^{\msp{\tau}}}\big)$. Evaluating at $\tau=T_i$ gives~\cref{eq:palive_var}.

\parhead{Derivation of~\cref{eq:post_pred_approx} ($q_\bphi(\mathcal{D}_i^\star\mid\mathcal{D}_i)$).}
The plug-in substitutes $q_\bphi(\bz_i\mid\mathcal{D}_i)$ for the exact posterior in $p_\btheta(\mathcal{D}_i^\star\mid\mathcal{D}_i)=\mathbb{E}[p(\mathcal{D}_i^\star\mid\bz_i)]$. A held-out history requires the customer to survive past their last held-out transaction, so conditioning on the event $\{\tau_i>t^\star_{i,y_i^\star}\}$,
\begin{equation*}
q_\bphi(\mathcal{D}_i^\star\mid\mathcal{D}_i) = q_\bphi\big(\tau_i>t^\star_{i,y_i^\star}\mid\mathcal{D}_i\big)\,\mathbb{E}_{q_\bphi(\bz_i\mid\mathcal{D}_i,\,\tau_i>t^\star_{i,y_i^\star})}\!\big[p(\mathcal{D}_i^\star\mid\bz_i)\big],
\end{equation*}
the survival factor from~\cref{eq:palive_var} and the likelihood~\cref{eq:indivlikelihood} evaluated on $\mathcal{D}_i^\star$ conditional on survival. This is~\cref{eq:post_pred_approx}. The conditional expectation is estimated by Monte Carlo~\cref{eq:postpred_mc}; we evaluate the bracketed Monte Carlo average in log space via the log-sum-exp trick, which keeps it numerically stable. 

\parhead{Evaluation of~\cref{eq:postpred_mc}.}
The per-draw held-out likelihood is the renewal complete-data likelihood under $\bz_i^{(s)}=(\tau_i,\lambda_i,k_i)$, with the first held-out inter-transaction time conditioned on no purchase having occurred between the last observed transaction and the end of the training window. Writing $f$ and $S$ for the ITT density and survival under $\bz_i^{(s)}$, and $\Delta t^\star_{i,1},\dots,\Delta t^\star_{i,y_i^\star}$ for the held-out ITTs, a draw with $\tau_i \geq T_i$ contributes
\begin{equation*}
p(\mathcal{D}_i^\star\mid\bz_i^{(s)}) = \frac{1}{S\!\left(T_i - t_{i,y_i}\right)}\left[\prod_{j=1}^{y_i^\star} f(\Delta t^\star_{i,j})\right] S\!\left(\min\left(T_i^\star,\,\tau_i\right)-t^\star_{i,y_i^\star}\right),
\end{equation*}
where the leading factor is the left-truncation correction---the reciprocal of the survival factor already included in the training likelihood of \cref{eq:indivlikelihood}---the product runs over held-out purchases, and the final survival term encodes no further purchase between the last held-out transaction and the earlier of the window end $T_i^\star$ and death $\tau_i$. For $y_i^\star=0$ the product is empty and $t^\star_{i,y_i^\star}=t_{i,y_i}$, so only the survival terms remain. A draw with $\tau_i < T_i$ is instead handled exactly: a customer dead before the end of the training window makes no held-out purchases with probability one, so $p(\mathcal{D}_i^\star\mid\bz_i^{(s)}) = \mathbb{1}(y_i^\star = 0)$. Under the truncated sampler of \cref{eq:postpred_mc} such draws arise only when $y_i^\star = 0$, since for $y_i^\star > 0$ every draw satisfies $\tau_i > t^\star_{i,y_i^\star} > T_i$.

\parhead{Derivation of~\cref{eq:forecast_approx_rt} (forecasted future frequency estimator).}
This is the posterior, finite-horizon analog of~\cref{eq:clv_rt}. Conditional on $\bz_i$ and survival to $T_i$, the residual process over $(T_i,\,T_i^\star\wedge\tau_i]$ accrues transactions at rate $\lambda_i$, so $\mathbb{E}[Y_i(T_i,T_i^\star\wedge\tau_i)\mid\bz_i]\approx\lambda_i\,(T_i^\star\wedge\tau_i-T_i)\,\mathbb{1}(\tau_i>T_i)$. Taking $\mathbb{E}_{q_\bphi}$ and using the mean-field factorization ($\lambda_i\perp\tau_i$ under $q_\bphi$),
\begin{align*}
\widehat{\mathbb{E}}^{\textrm{RT}}_{\bphi}\big[Y_i(T_i,T_i^\star)\mid\mathcal{D}_i\big] &= \mathbb{E}_{q_\bphi}[\lambda_i]\,\mathbb{E}_{q_\bphi}\!\big[(T_i^\star\wedge\tau_i-T_i)\,\mathbb{1}(\tau_i>T_i)\big] \\
&= \mathbb{E}_{q_\bphi}[\lambda_i]\,q_\bphi(\textsc{alive}\mid\mathcal{D}_i)\,\mathbb{E}_{q_\bphi}\!\big[T_i^\star\wedge\tau_i-T_i\mid\tau_i>T_i\big],
\end{align*}
which is~\cref{eq:forecast_approx_rt}; the $k_i$-marginal drops because $\lambda_i$ alone sets the rate. The rate factor $\mathbb{E}_{q_\bphi}[\lambda_i]$ and the survival factor $q_\bphi(\textsc{alive}\mid\mathcal{D}_i)$ are elementary Weibull moments of the variational marginals. The residual-lifetime factor $\mathbb{E}_{q_\bphi}[T_i^\star\wedge\tau_i-T_i\mid\tau_i>T_i]$ is not an elementary moment, but a regularized incomplete-gamma expression in closed form: writing $(\tilde s,\tilde r)$ for $(\tilde s_i^{\msp{\tau}},\tilde r_i^{\msp{\tau}})$ and $P$ for the regularized lower incomplete gamma function, it equals $\tfrac{\Gamma(1+\nicefrac{1}{\tilde s})}{\tilde r}\big[P\big(\nicefrac{1}{\tilde s},\,[\tilde r(T_i^\star-t_{i,y_i})]^{\tilde s}\big)-P\big(\nicefrac{1}{\tilde s},\,[\tilde r(T_i-t_{i,y_i})]^{\tilde s}\big)\big]\big/\,q_\bphi(\textsc{alive}\mid\mathcal{D}_i)$. 

\parhead{Proof of~\Cref{prop:rt_bias} (error bound of forecast estimator).}
Since $-1\le \mathbb{E}[Y_i(t)\mid\bz_i]-\lambda_i t\le c(k_i)-1$ for all $t$ (elementary-renewal lower bound and Lorden upper bound), the count over $(T_i,\,T_i^\star\wedge\tau_i]$ deviates from its rate ($\lambda_i$) value by at most the band width $c(k_i)$: conditional on $\tau_i>T_i$, $\big|\mathbb{E}[Y_i(T_i,T_i^\star\wedge\tau_i)\mid\bz_i]-\lambda_i(T_i^\star\wedge\tau_i-T_i)\big|\le c(k_i)$, with $c$ as in~\Cref{prop:clv_bias}, and $0$ when $\tau_i\le T_i$. Taking $\mathbb{E}_{q_\bphi}$ and factorizing the survival event under the mean-field $q_\bphi$ ($k_i\perp\tau_i$) yields~\looseness=-1
\begin{equation*}
\big|\widehat{\mathbb{E}}^{\textrm{RT}}_{\bphi}\big[Y_i(T_i,T_i^\star)\mid\mathcal{D}_i\big]-\mathbb{E}_{\bphi}\big[Y_i(T_i,T_i^\star)\mid\mathcal{D}_i\big]\big| \le q_\bphi(\textsc{alive}\mid\mathcal{D}_i)\,\mathbb{E}_{q_\bphi(k_i\mid\mathcal{D}_i)}\!\big[c(k_i)\big],
\end{equation*}
which is~\cref{eq:rt_bias_bound}.

\section{Algorithmic and implementational details}
\label{sec:app_details}

\parhead{Recognition network.} The summary $u(\mathcal{D}_i)$ stacks $(y_i,\,t_{i,y_i},\,\kappa_i,\,T_i)$ with the moment features $\zeta^{\msp{p}}_i=\sum_j(\itt_{i,j})^{\nicefrac{p}{10}}$ for $p=1,\dots,P$, with $P=25$ in all runs. The recognition network $\bphi(\cdot)$ is built from an MLP with two-to-three hidden layers (widths $256\text{--}64\text{--}32$ for the synthetic studies, $(8,8)$ for the sub-model study of \Cref{subsec:submodels}, and $256\text{--}128\text{--}64$ for the full-panel fit) and ReLU activations; its six outputs are mapped to positive shapes and rates and clamped to a numerically stable range. Concretely, the recognition network maps the summary to variational parameters as $\bphi_i = \exp\big(\bg + f_{\bpsi}(\bu_i)\big)$, where $f_{\bpsi}$ is the MLP with weights $\bpsi$ and $\bg\in\mathbb{R}^6$ is a shared, per-dimension global offset applied in log-space---the shared global variational parameters. The full variational parameter set is $\bphi \defeqinl \{\bpsi,\bg\}$.

\parhead{Optimizer and learning-rate schedule.} We use Adam/AdamW with two parameter groups---the recognition-network weights, and everything global (the shared global variational parameters together with the prior hyperparameters)---each with its own weight decay ($10^{-4}$ on the network weights, $0$ on the global group) and, in principle, its own step size, though all reported runs share one initial rate. The learning rate follows a cosine schedule to near zero with no warmup and no restarts; initial rates are $3\times10^{-4}$ (synthetic studies; $10^{-3}$ for the sub-model study) and $5\times10^{-5}$ (full-panel fit).

\parhead{Gradient stability.} Reparameterization noise is the truncated exponential $\varepsilon=-\log u$ with $u\sim\textrm{Uniform}(0.01,0.99)$ (\Cref{subsec:details}), and gradients are additionally clipped to global norm $10$ at every step. 

\parhead{Inference compilation and early stopping.} The recognition network is pretrained on $5{,}000$ ($20{,}000$ for the full panel) synthetic customers for $500$ ($1{,}000$) Adam steps. When early stopping is enabled we hold out $20\%$ of customers, stratified by test-window activity with a fixed seed, and stop when their held-out predictive log-likelihood has not improved for $100$ evaluations, restoring the best checkpoint; population-parameter recovery (\Cref{subsec:param_recovery}) and the full-panel fit instead run a fixed iteration budget. 

\parhead{Confidence intervals for $\widehat{\btheta}$ in \Cref{subsec:param_recovery}.} We computed $95\%$ confidence intervals for WTYD's hyperparameter estimate $\widehat{\btheta}$ in two ways, both from the single observed dataset rather than across simulation replicates. The first is a customer-level bootstrap: we resampled the $n$ customers with replacement, refit WTYD from scratch on the resampled panel with the same settings and iteration budget as the original fit, repeated this $100$ times, and took the $2.5$th and $97.5$th percentiles of the resulting estimates. The second is an asymptotic approximation. Treating $\widehat{\btheta}$ as an M-estimator, i.e., as the maximizer of the ELBO with the recognition network held fixed at its fitted value $\widehat{\bphi}$, we estimated its covariance by the inverse Hessian
\begin{equation*}
\widehat{\textrm{Cov}}(\widehat{\btheta}) \,=\, \Big[ -\nabla^2_{\btheta}\, \mathcal{B}(\widehat{\bphi}, \btheta) \,\Big|_{\btheta = \widehat{\btheta}} \Big]^{-1}
\end{equation*}
and formed intervals on the log scale to respect positivity. Across the $18$ regime-hyperparameter combinations of \Cref{fig:param_recovery_hetT}, the asymptotic standard error is between $0.40$ and $0.80$ times the bootstrap standard deviation, with the greatest shortfall on the regular-regime $k$ hyperparameters. The asymptotic approximation conditions on the fitted recognition network and uses the curvature of the ELBO in place of that of the marginal likelihood, and both simplifications plausibly understate variability. We therefore recommend the bootstrap whenever calibrated intervals matter. Pareto/GGG's interval in \Cref{fig:param_recovery_hetT} is its native $95\%$ equal-tailed posterior credible interval, computed from the same MCMC draws as its point estimate.

\parhead{Software and reproducibility.} The reference implementation is in \textsc{PyTorch}~\citep{ansel2024pytorch2}, with all subsampling, reparameterization noise, and Monte Carlo sampling driven by a single fixed seed ($42$) per run. One estimator caveat: Pareto/GGG's held-out means average over customers with a finite Monte Carlo estimate. For roughly $1\%$ of customers per subset, every retained MCMC draw places $\tau$ before the customer's last held-out transaction, so the finite-sample estimate is zero probability ($-\infty$ log-likelihood) even though the exact model probability is positive; these customers are excluded from Pareto/GGG's mean (their count is reported alongside each run's outputs), while WTYD's means average all customers. A second implementation note: WTYD's training loop skips any iteration whose minibatch loss is non-finite rather than halting. Instrumented re-runs of eight representative fits (four of the $n{=}1{,}000$ retail subsets and four contrasting synthetic $n{=}10{,}000$ cells) measured how often this fires: never in five of the eight fits, and between $0.02\%$ and $0.22\%$ of iterations in the remaining three (at most $7$ iterations in any single fit, concentrated in heavy-tailed and clumpy regimes); the companion guard against runtime errors never fired at all. Skip counts are logged with every fit. \Cref{tab:app_experiment_settings} shows the per-experiment settings for each figure in the simulation studies. Note that, for the second study, held-out data is generated but unused, and the clumpy regime is additionally swept to $3.2{\times}10^5$ iterations to check that its $s^{\msp{k}}$ underestimate is not a training-budget artifact. The figures for the third and fourth studies report the $n{=}10{,}000$ datasets.~\looseness=-1 

\begin{table}[t]
\centering
\footnotesize
\setlength{\tabcolsep}{4pt}
\caption{Per-experiment settings for the four simulation studies (\Cref{sec:performance_eval}). Abbreviations: Ga $=$ Gamma, Pa $=$ Pareto, U $=$ Uniform, Wei $=$ Weibull. All WTYD fits use $P=25$ power features and seed $42$; the first study differs by design (a conditioned sub-model with observed $\tau$ and $k\equiv1$, hence a smaller network, distinct optimizer settings, and no Pareto/GGG comparison). The $k$-regime Uniform priors are stated in Gamma-shape units; Weibull cases use the matched Weibull-$k$ equivalents (\Cref{prop:correspondance_prop}); ``mixture'' denotes an equal mixture of the two Uniforms and a point mass at $k=1$. $^\dagger$Bootstrap refits use $(0.01,0.99)$; the point-estimate fit's noise bounds were not separately logged. Reference implementation in \textsc{PyTorch}~\citep{ansel2024pytorch2}; Pareto/GGG via \textsc{BTYDplus} in \textsc{R}.}
\label{tab:app_experiment_settings}
\begin{tabular}{l cccc}
\toprule
 & \textbf{Posterior} & \textbf{Parameter} & \textbf{Latent} & \textbf{Forecast} \\
 & \textbf{comparison} & \textbf{recovery} & \textbf{recovery} & \textbf{quality} \\
 & (\Cref{fig:submodels}) & (\Cref{fig:param_recovery_hetT}) & (\Cref{fig:rank_corr_n10000_simulation_results}) & (\Cref{fig:higher_n_simulation_results}) \\
\midrule
\multicolumn{5}{l}{\emph{Data generation}}\\
$n$                    & $1000$ & $1000$ & $1000,10000$ & $1000,10000$ \\
seeds                  & $1$    & $1$    & $6$          & $6$ \\
window $T$             & $1$    & $\mathrm{U}(1,10)$ & $1$ & $1$ \\
holdout $(T^\star{-}T)$& --     & $0.5$ & $0.5$ & $0.5$ \\
$\lambda$ prior        & $\mathrm{Ga}(10,1)$ & $\mathrm{Ga}(2,0.2)$ & $\mathrm{Ga}(4,0.15)$ & $\mathrm{Ga}(4,0.15)$ \\
$\tau$ prior           & $\mathrm{Pa}(2,5)$ & $\mathrm{Pa}(2,5)$ & \makecell{$\mathrm{Pa}(2,2.5)$\\$\mathrm{Pa}(2,8.5)$} & \makecell{$\mathrm{Pa}(2,2.5)$\\$\mathrm{Pa}(2,8.5)$} \\
$k$ prior              & $k\equiv1$ & \makecell{$\mathrm{Ga}(25,10)$\\$\mathrm{Ga}(65,80)$\\$\mathrm{Ga}(7,4)$} & \makecell{$\mathrm{U}(1.5,10)$\\$\mathrm{U}(0.5,0.95)$\\mixture} & \makecell{$\mathrm{U}(1.5,10)$\\$\mathrm{U}(0.5,0.95)$\\mixture} \\
renewal                & exp.\ ($k{=}1$) & \makecell{Ga \& Wei\\(matched)} & Ga, Wei & Ga, Wei \\
churn axis             & --     & --     & high, low & high, low \\
\# datasets            & $1$    & $6$    & $144$     & $144$ \\
\midrule
\multicolumn{5}{l}{\emph{WTYD fit}}\\
optimizer              & Adam   & AdamW  & AdamW & AdamW \\
lr (both groups)       & $10^{-3}$ & $3{\times}10^{-4}$ & $3{\times}10^{-4}$ & $3{\times}10^{-4}$ \\
weight decay (net/$\btheta$) & -- & $10^{-4}/0$ & $10^{-4}/0$ & $10^{-4}/0$ \\
iterations             & $10^4$ & $8{\times}10^4$ & $2{\times}10^4$ & $2{\times}10^4$ \\
$\text{shape}_{\max}$  & $10$   & $80$   & $10$  & $10$ \\
minibatch              & full   & $200$  & $200$ & $200$ \\
val.\ frac / patience  & --     & $0$ / off & $0.2/100$ & $0.2/100$ \\
pretrain (samp/iter)   & --     & $5000/500$ & $5000/500$ & $5000/500$ \\
MC draws (tr/final)    & --     & $10/1$ & $100/200$ & $100/200$ \\
recog.\ MLP            & $(8,8)$ & $256$-$64$-$32$ & $256$-$64$-$32$ & $256$-$64$-$32$ \\
noise $u$-range        & $(0.01,0.99)$ & $(0.01,0.99)^\dagger$ & $(0.01,0.99)$ & $(0.01,0.99)$ \\
\bottomrule
\end{tabular}
\end{table}